\documentclass[aps,pra,reprint,superscriptaddress,longbibliography]{revtex4-2}
\usepackage{graphicx}
\usepackage{mathtools}
\usepackage{amssymb}
\usepackage{bm}
\usepackage{braket}
\usepackage{xcolor}
\usepackage[colorlinks=true,linkcolor=blue,citecolor=magenta,urlcolor=blue,filecolor=blue]{hyperref}

\DeclareMathAlphabet{\mathcal}{OMS}{cmsy}{m}{n}

\newcommand{\Tr}[1]{{\operatorname{Tr}\left[#1\right]}}

\newlength{\ketketwidth}
\newlength{\ketwidth}

\newcommand{\kettstylesep}[3]{
 \settowidth{\ketwidth}{$#2\left|#1\right\rangle$}
 \settowidth{\ketketwidth}{$#2\left.\left|#1\right\rangle\right\rangle$}
 \left|#1\right\rangle#3\hspace{\ketwidth}\hspace{-\ketketwidth}
}

\newcommand{\kett}[1]{
 \left.\mathchoice
 {\kettstylesep{#1}{\displaystyle}{\hspace{0.3em}}}
 {\kettstylesep{#1}{\textstyle}{\hspace{0.3em}}}
 {\kettstylesep{#1}{\scriptstyle}{\hspace{0.3em}}}
 {\kettstylesep{#1}{\scriptscriptstyle}{\hspace{0.25em}}}
 \right\rangle
}

\newcommand{\bbrastylesep}[3]{
 \settowidth{\ketwidth}{$#2\left\langle#1\right|$}
 \settowidth{\ketketwidth}{$#2\left\langle\left\langle#1\right|\right.$}
 #3\hspace{\ketwidth}\hspace{-\ketketwidth}\left\langle#1\right|
}
\newcommand{\bbra}[1]{
 \left\langle\mathchoice
 {\bbrastylesep{#1}{\displaystyle}{\hspace{0.3em}}}
 {\bbrastylesep{#1}{\textstyle}{\hspace{0.3em}}}
 {\bbrastylesep{#1}{\scriptstyle}{\hspace{0.3em}}}
 {\bbrastylesep{#1}{\scriptscriptstyle}{\hspace{0.25em}}}
 \right.
}

\newcommand{\kettbbra}[2]{\kett{#1}\hspace{-0.2em}\bbra{#2}}

\newcommand{\bbrakettstylesep}[4]{
 \settowidth{\ketwidth}{$#3\left\langle#1\middle|#2\right\rangle$}
 \settowidth{\ketketwidth}{$#3\left\langle\left\langle#1\middle|#2\right\rangle\right.$}
 #4\hspace{\ketwidth}\hspace{-\ketketwidth}\left\langle#1\middle|#2\right\rangle#4\hspace{\ketwidth}\hspace{-\ketketwidth}
}
\newcommand{\bbrakett}[2]{
 \left\langle\mathchoice
 {\bbrakettstylesep{#1}{#2}{\displaystyle}{\hspace{0.3em}}}
 {\bbrakettstylesep{#1}{#2}{\textstyle}{\hspace{0.3em}}}
 {\bbrakettstylesep{#1}{#2}{\scriptstyle}{\hspace{0.3em}}}
 {\bbrakettstylesep{#1}{#2}{\scriptscriptstyle}{\hspace{0.25em}}}
 \right\rangle
}

\begin{document}

\title{Response kinetic uncertainty relation for Markovian open quantum systems}

\author{Kangqiao Liu}
\email{kqliu@xhu.edu.cn}
\affiliation{School of Science, Key Laboratory of High Performance Scientific Computation, Xihua University, Chengdu 610039, China}

\author{Jie Gu}
\email{jiegu1989@gmail.com}
\affiliation{Chengdu Academy of Educational Sciences, Chengdu 610036, China}

\begin{abstract}
Response uncertainty relations in stochastic thermodynamics extend precision bounds to the sensitivity of observables under external perturbations.
Here we derive a quantum response kinetic uncertainty relation for continuously monitored Markovian open quantum systems in the steady state of the Lindblad master equation.
The response precision of a measured trajectory observable is bounded by two contributions: the conventional quantum dynamical activity and a perturbation-induced intersubspace transition term.
The latter is absent in the classical limit and captures a genuinely quantum part of the response cost.
We identify simple conditions under which either contribution vanishes, and we further clarify the structure of the intersubspace term through a symmetry-resolved decomposition and exact sector-selection rules.
The bound and its structure are illustrated in a driven two-level atom.
\end{abstract}

\maketitle

\section{Introduction}\label{sec:introduction}
Thermodynamic uncertainty relations (TUR) \cite{barato2015,horowitz2020,liu2020tur,Dieball2023} and kinetic uncertainty relations (KUR) \cite{diterlizzi2019,Yan2019,macieszczak2024,PrechLandi2025,Moreira2025} provide fundamental limits to the precision of a nonequilibrium observable in terms of the thermodynamic or kinetic properties of the system.
Various efforts have endeavored to generalize the TUR and the KUR to Markovian open quantum systems \cite{hasegawa2020,Kewming2024,prech2024}, which set fundamental trade-off relationships between the precision of an observable and the thermodynamic or kinetic properties of those systems.
The stochastic observables, for example, the current, that appear in these bounds are usually measured through continuous monitoring of the quantum system, namely through the unraveling of the Lindblad master equation in quantum trajectory theory \cite{gong2016,liuCharacteristic2016,carollo2019,lidar2020}.
Since the information of the measured quantum system is obtained through stochastic time series of quantum jumps, the precision of the measured time series is of crucial importance \cite{He2023,landi2024}.
Additionally, continuous quantum measurement itself plays an important role both theoretically and practically in many fields, including quantum sensing and quantum metrology \cite{Rossi2020,Ilias2022,Yang2023}, atomic-molecular-optical physics \cite{ueda1990,bakr2009,sherson2010,Patil2015,Ashida2015,Wigley2016,gross2021}, condensed-matter physics \cite{Fuji2020,Jian2021,muller2022,Krishna2023,yokomizo2024}, and thermodynamics \cite{Ashida2018,Kewming2022entropyproduction,Manzano2022,Das2023}.

Recently, response uncertainty relations have extended the fluctuation problem to static responses under external perturbations in classical Markov jump processes.
For classical systems, both the dissipation-controlled response TUR and the activity-controlled response KUR (R-KUR) are now available, and subsequent works have clarified their relation through exact fluctuation-response relations and more general fluctuation-response inequalities that place kinetic and entropic perturbations within a unified Cram\'er--Rao framework for broader classes of observables and finite observation times \cite{ptaszynski2024,liu2024,aslyamov2024,kwon2024fri}.
The classical R-KUR can be written explicitly as \cite{liu2024}
\begin{equation}
\label{eq:classical_rkur}
\frac{(\partial_{\theta}\langle O \rangle)^2}{\langle\langle O \rangle\rangle} \le a_{\rm max}^2 \mathcal{A}_{\rm cl}.
\end{equation}
where $O$ is a stochastic observable with a mean $\langle O \rangle$ and a variance $\langle\langle O \rangle\rangle$ with respect to a path probability parametrized by $\theta$, $a_{\rm max}^2$ is called the perturbative rate, and $\mathcal{A}_{\rm cl}:= \int_0^T {\rm d}t \sum_{i\ne j} p_i W_{ij}/T$ is the time-averaged dynamical activity, abbreviated as DA, for a classical Markov jump process and characterizes the frequency of stochastic transitions.

For Markovian open quantum systems, the fluctuation side has developed from quantum TUR and KUR to broader trade-off relations for quantum trajectory observables \cite{hasegawa2020,NakajimaUtsumi2023,Kewming2024,prech2024,vanvu2024}.
On the response side, recent works have established activity-centered quantum response bounds and quantum fluctuation-response inequalities for Lindblad dynamics \cite{vanvu2024,kwon2024fri}.
A steady-state quantum R-KUR (QR-KUR) for continuously monitored Lindblad observables with general parameter encoding remains to be formulated in a form that closes directly with quantum activity and identifies the additional genuinely quantum contribution on the right-hand side.

In this work, we generalize the classical R-KUR \eqref{eq:classical_rkur} to the steady state of an open quantum system obeying the Lindblad dynamics.
The resulting response bound contains two contributions: one proportional to the conventional quantum DA, in direct analogy with the classical R-KUR, and one determined by perturbation-induced quantum transitions.
Sufficient conditions under which either contribution vanishes are given and the bound is illustrated in a driven two-level model.
We also illustrate the structure of this genuinely quantum contribution by answering which decaying sectors can contribute to it, and under what symmetry conditions it is forbidden or restricted. We address this question through a group-theoretic, symmetry-resolved decomposition of the intertransition term. Finally, the issue of non-unique steady states is discussed.

\section{Monitored GKSL setting}\label{sec:setup}
The dynamics of a finite-dimensional Markovian open quantum system $S$ with Hamiltonian $H(\theta)$ can be described by the Gorini-Kossakowski-Sudarshan-Lindblad (GKSL) form master equation, or in short, Lindblad master equation \cite{Lindblad1976}. Examples include photon detection \cite{ueda1990} and temperature estimation with thermometry \cite{boeyens2023}:
\begin{equation}\label{eq:lindblad_eq}
 \dot{\rho}(t)=\mathcal{L}(\theta)\rho(t),
\end{equation}
where the time-independent Lindbladian $\mathcal{L}(\theta)$ is a superoperator defined as
\begin{equation}\label{eq:Lindbladian}
 \mathcal{L}(\theta)\rho:= -i[H(\theta), \rho]+\sum_c \mathcal{D}[L_c(\theta)]\rho,
\end{equation}
where the parameter $\theta$ is encoded in the system Hamiltonian and/or in the jump operators $L_c(\theta)$, and the dissipator is defined as $\mathcal{D}[L]\rho := L\rho L^{\dagger} - \{L^{\dagger}L,\rho\}/2$.
Throughout this work we set $\hbar=1$.
The $\theta$-dependence will be dropped in the following to simplify the notation unless necessary.
Throughout the formal derivation, $\theta$ denotes a generic perturbation parameter; in concrete examples it is instantiated by physical controls such as $\gamma$, $\Omega$, $\beta$, $\omega_d$, or $\omega$.

An empirical observable is measured by the unraveling of the Lindblad equation \cite{carollo2019}.
The number of quantum jumps is recorded during a time interval $[0,T]$ as ${\rm d}N_c(t)=1$ if a quantum jump in channel $c$ is observed at time $t$; otherwise ${\rm d}N_c(t)=0$.
The probability of observing a jump is given by $\Tr{L_c^{\dagger}L_c\rho(t)}{\rm d}t$ and that of a no-jump event is $1-\sum_c\Tr{L_c^{\dagger}L_c\rho(t)}{\rm d}t$ \cite{gong2016}.
The observable is calculated according to the record as $\Phi(T):=\sum_c \nu_c \int_0^T {\rm d}N_c(t)$ where $\nu_c$ depends on the physical observable of interest.
For example, $\Phi(T)$ with $\nu_c=\pm 1$ is the net particle current, whereas $\Phi(T)$ with $\nu_c=1$ is the total number of quantum jumps.
The rate of $\Phi(T)$ at time $t$ is defined as $\phi(t):={\rm d}\Phi(t)/{\rm d}t=\sum_c \nu_c {\rm d}N_c(t)/{\rm d}t$.
In the steady state, we denote its long-time average rate simply by $\langle \phi \rangle$.

\section{Quantum response kinetic uncertainty relation}\label{sec:qrkur}
\subsection{Main result}\label{subsec:main_result}
In the long-time limit $T\to\infty$, the system $S$ evolves to its steady state $\pi$.
In Appendix \ref{app:proof_qrkur}, we prove that the response in the average rate $\phi$ of an observable to a perturbation in the parameter $\theta$ obeys a QR-KUR:
\begin{equation}
\label{eq:lindblad_rkur}
\frac{(\partial_\theta \langle \phi \rangle)^2}{\langle \langle \phi \rangle \rangle}
\le \mathsf{a}_{\max}^2 \mathcal{A} + \mathcal{Z}.
\end{equation}
where
\begin{equation}
\mathsf{a}_{\max}^2:=4\max_c\frac{\Tr{\partial_\theta L_c\pi\partial_\theta L_c^{\dagger}}}{\Tr{L_c\pi L_c^{\dagger}}}
\end{equation}
and
\begin{equation}\label{eq:DA}
 \mathcal{A}:=\sum_{c}\Tr{L_c\pi L_c^\dagger}
\end{equation}
is the conventional quantum DA which by definition counts the frequency of quantum jumps in all channels \cite{hasegawa2020,Kewming2024,landi2024}. The quantity $\mathcal Z$ is a perturbation-induced intersubspace transition term defined by
\begin{equation}
\label{eq:Z_definition}
\mathcal Z:=-8{\rm Re}\left[
\sum_{j\ne 0} \frac{1}{\lambda_j} \bbra{\mathbf{1}} \hat{\mathcal{K}}_1 \kettbbra{x_j}{y_j}
\hat{\mathcal{K}}_2 \kett{\pi}
\right].
\end{equation}
Vectorization is adopted to map superoperators to matrices and density operators to vectors, $\lambda_j\in\mathbb{C}$ are the eigenvalues of the matrix representation of the Lindbladian, and $\kett{x_j}$ and $\bbra{y_j}$ are the corresponding right- and left-eigenvectors. In particular, for $\lambda_0=0$, $\kett{x_0}=\kett{\pi}$ is the vectorized steady-state density operator and $\bbra{y_0}=\bbra{\mathbf{1}}$ the vectorized identity matrix.
The corresponding perturbation superoperators are
\begin{align}
& \mathcal{K}_1 \rho := -i (\partial_{\theta}H_{\rm eff}) \rho + \sum_c \left(\partial_{\theta}L_c\right) \rho L_c^{\dagger},\\
& \mathcal{K}_2 \rho := i\rho \partial_{\theta} H^{\dagger}_{\rm eff} + \sum_c L_c \rho \partial_{\theta}L_c^{\dagger}.
\end{align}
with the non-Hermitian effective Hamiltonian $H_{\rm eff}:= H - i\sum_c L_c^{\dagger}L_c/2$.
The intertransition contribution is not guaranteed to be non-negative as shown in Appendix~\ref{app:signofZ}.

\begin{figure}[t]
 \centering
 \includegraphics[width=\columnwidth]{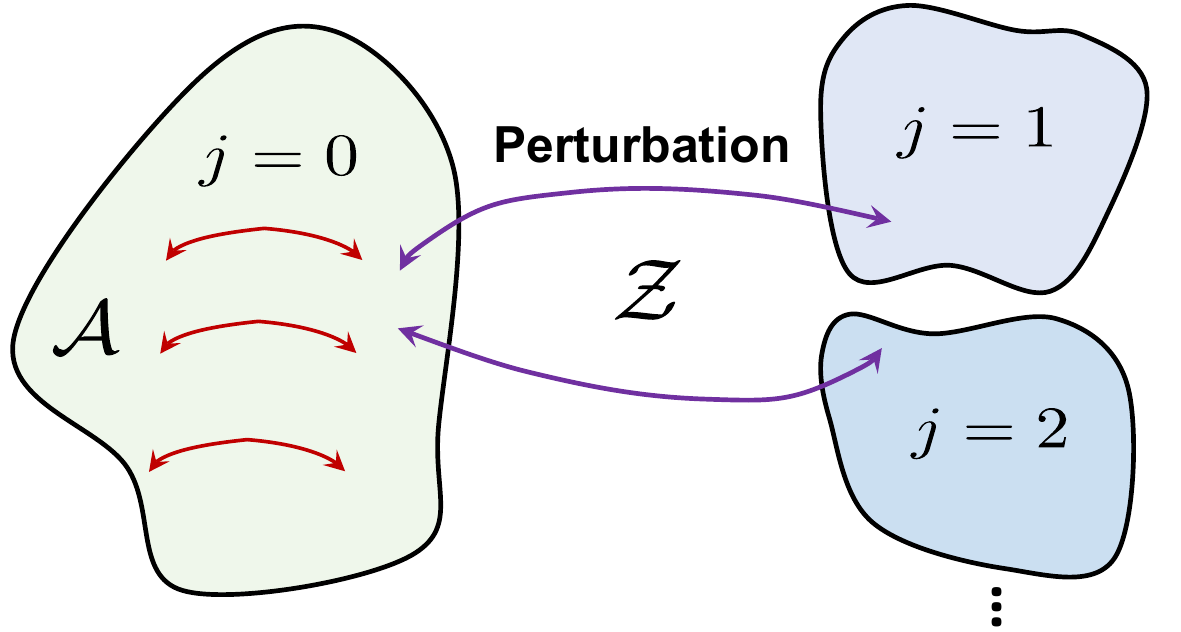}
 \caption{Schematic illustration of the right-hand side of the QR-KUR. The left green area represents the steady-state sector with $j=0$. The red arrows indicate quantum jumps within the steady-state sector, contributing to the quantum dynamical activity $\mathcal A$. The two blue areas represent two instances of the complementary sectors, and the purple arrows represent the perturbation-induced couplings between the steady-state sector and its complementary sectors, yielding the intertransition term $\mathcal Z$. No transitions entirely within one complementary sector or between two complementary sectors contribute to the right-hand side.}
 \label{fig:schematic_traffic}
\end{figure}

\subsection{Interpretation}\label{subsec:interpretation_rhs}
The two contributions on the right-hand side of inequality~\eqref{eq:lindblad_rkur} play distinct roles.
The term $\mathsf{a}_{\max}^2\mathcal{A}$ is built from the conventional quantum DA and is determined by quantum jumps within the steady-state sector.
By contrast, $\mathcal Z$ is nonzero only when the perturbation characterized by $\mathcal K_{1,2}$ couples the steady-state sector to at least one decaying sector.
Figure~\ref{fig:schematic_traffic} schematically illustrates this separation: $\mathcal A$ counts intra-steady-sector transitions, whereas $\mathcal Z$ accounts for perturbation-induced couplings between the steady-state sector and its complementary sectors.

This distinction is transparent in the classical limit such that $L_{ji}=\sqrt{W_{ji}}|\epsilon_j\rangle\langle\epsilon_i|$, where $\{|\epsilon_i\rangle\}$ is the eigenbasis, and the off-diagonal elements of the steady-state density matrix vanish, $\pi_{ij}=0$ for $i\ne j$.
The quantum DA \eqref{eq:DA} reduces to the classical one $\mathcal{A}_{\rm cl}=\sum_{i\ne j}\pi_{ii}W_{ji}$ in the steady state and
\begin{equation}
\mathsf{a}_{\max}^2=4\max_{i\ne j}\frac{(\partial_\theta\sqrt{W_{ij}})^2}{W_{ij}}=\max_{i\ne j}(\partial_\theta\ln W_{ij})^2,
\end{equation}
which also reduces to the perturbative rate $a_{\max}^2$ in the classical R-KUR \eqref{eq:classical_rkur}.
The intertransition term vanishes in the classical limit. In that case the trace pairings entering $\mathcal Z$ vanish identically, and only the activity term remains.
Therefore, the term $\mathsf{a}_{\max}^2\mathcal A$ in inequality~\eqref{eq:lindblad_rkur} is a quantum generalization of the activity term in the classical R-KUR, while $\mathcal Z$ is genuinely quantum.
There is no contribution from transitions entirely within one complementary sector or between two complementary sectors.
The steady-state fluctuations and response are therefore constrained only by transitions tied to the steady-state sector and by perturbation-induced excursions out of it.

We consider two special regimes.
If the perturbed parameter $\theta$ is encoded only in the Hamiltonian and not in the jump operators $L_c$, then $\mathsf a_{\max}^2=0$.
The QR-KUR remains nontrivial because the right-hand side reduces to the intertransition contribution $\mathcal Z$.
If the perturbed parameter $\theta$ is not encoded in the Hamiltonian and the jump operators take the multiplicative form
\begin{equation}
\label{eq:multiplicative_jump_form_main}
L_c(\theta)=g_c(\theta)L_c,\quad g_c(\theta)\in\mathbb R.
\end{equation}
with $L_c$ independent of $\theta$, then $\mathcal Z=0$ by the same trace-cyclicity mechanism.
In that case the perturbative rate reduces to
\begin{equation}
\mathsf a_{\max}^2=4\max_c(\partial_\theta\ln g_c)^2,
\end{equation}
which is the direct quantum analogue of the classical perturbative rate in inequality~\eqref{eq:classical_rkur}.

Recent quantum developments include finite-time fluctuation-response inequalities for monitored open systems \cite{kwon2024fri}, activity-controlled QR-KURs for general observables \cite{vanvu2024}, and quantum KUR analyses in which coherence modifies the current-noise trade-off itself \cite{prech2024}.
The steady-state Lindblad setting considered here yields a response bound whose right-hand side separates into the activity contribution $\mathsf{a}_{\max}^2\mathcal A$ and the intersector term $\mathcal Z$.
For purely multiplicative jump-amplitude perturbations one has $\mathcal Z=0$, and the bound reduces to the activity-only structure.
In the parametrization $H(\theta)=(1+\theta)H$ and $L_c(\theta)=\sqrt{1+\theta}L_c$, the present result reduces at $\theta\to 0$ to Hasegawa's quantum KUR \cite{hasegawa2020}.

\section{Driven two-level atom}\label{sec:two_level}
The QR-KUR \eqref{eq:lindblad_rkur} is illustrated in a driven two-level atom under dissipation described by the Hamiltonian \cite{gammelmark2014,hasegawa2020,NakajimaUtsumi2023,landi2024}
\begin{equation}
H = \frac{\Delta}{2}\sigma_z + \Omega \sigma_x.
\end{equation}
where $\Delta:=\omega-\omega_d$ is the detuning between the transition frequency $\omega$ of the two-level atom and the frequency $\omega_d$ of the driving laser, and $\Omega$ is the Rabi frequency.
The atom is coupled to a heat bath at inverse temperature $\beta$.
The jump operators are $L_-:=\sqrt{\gamma(\bar N+1)}\sigma_-$ and $L_+:=\sqrt{\gamma\bar N}\sigma_+$, where $\bar N:=1/[\exp(\beta\omega)-1]$ is the Bose--Einstein occupation and $\gamma$ is the dissipation strength.

It is convenient to work in the column-stacking basis
$\{\ket{e}\bra{e},\ket{g}\bra{e},\ket{e}\bra{g},\ket{g}\bra{g}\}$,
in which the Liouvillian matrix is
\begin{equation}
\hat{\mathcal L}=
\begin{pmatrix}
-\gamma(\bar N+1) & -i\Omega & i\Omega & \gamma\bar N \\
-i\Omega & i\Delta-\frac{\gamma(2\bar N+1)}{2} & 0 & i\Omega \\
 i\Omega & 0 & -i\Delta-\frac{\gamma(2\bar N+1)}{2} & -i\Omega \\
 \gamma(\bar N+1) & i\Omega & -i\Omega & -\gamma\bar N
\end{pmatrix}.
\label{eq:L_two_level_general_methods}
\end{equation}
The observable superoperators entering the mean and variance are
\begin{align}
& \hat{\mathcal J}=\nu_-\gamma(\bar N+1)\sigma_-^*\otimes\sigma_-+\nu_+\gamma\bar N\sigma_+^*\otimes\sigma_+,\\
& \hat{\mathcal P}=\nu_-^2\gamma(\bar N+1)\sigma_-^*\otimes\sigma_-+\nu_+^2\gamma\bar N\sigma_+^*\otimes\sigma_+.
\end{align}
so that
\begin{align}
& \langle\phi\rangle=\bbra{\mathbf 1}\hat{\mathcal J}\kett{\pi},\  \langle \langle\phi\rangle \rangle=\mathcal P_\phi-2\bbra{\mathbf 1}\hat{\mathcal J}\hat{\mathcal L}^{\mathrm D}\hat{\mathcal J}\kett{\pi},\\
& \mathcal P_\phi:=\bbra{\mathbf 1}\hat{\mathcal P}\kett{\pi}.
\end{align}
We denote by
$\hat{\mathcal L}^{\mathrm D}:=\sum_{j\ne 0}\lambda_j^{-1}\kettbbra{x_j}{y_j}$
the Drazin pseudoinverse of $\hat{\mathcal L}$ on the decaying subspace.

The steady state is
{\footnotesize
\begin{align}
\pi=\frac{1}{(1+2\bar N)\alpha}
\begin{pmatrix}
\bar N\alpha+4\Omega^2 & -2\Omega\bigl(2\Delta+i\gamma(1+2\bar N)\bigr)\\
-2\Omega\bigl(2\Delta-i\gamma(1+2\bar N)\bigr) & (1+\bar N)\alpha-4\Omega^2
\end{pmatrix},
\label{eq:pi_two_level_general_methods}
\end{align}
}
where $\alpha:=4\Delta^2+\gamma^2(1+2\bar N)^2+8\Omega^2$, and the corresponding current and activity are
\begin{equation}
\label{eq:JA_two_level_general_methods}
\langle J\rangle=-\frac{4\gamma\Omega^2}{\alpha},
\quad
\mathcal A=2\gamma\bar N\frac{1+\bar N}{1+2\bar N}+\frac{4\gamma\Omega^2}{(1+2\bar N)\alpha}.
\end{equation}

\subsection{Limiting case}

\begin{figure}[t]
 \centering
 \includegraphics[width=\columnwidth]{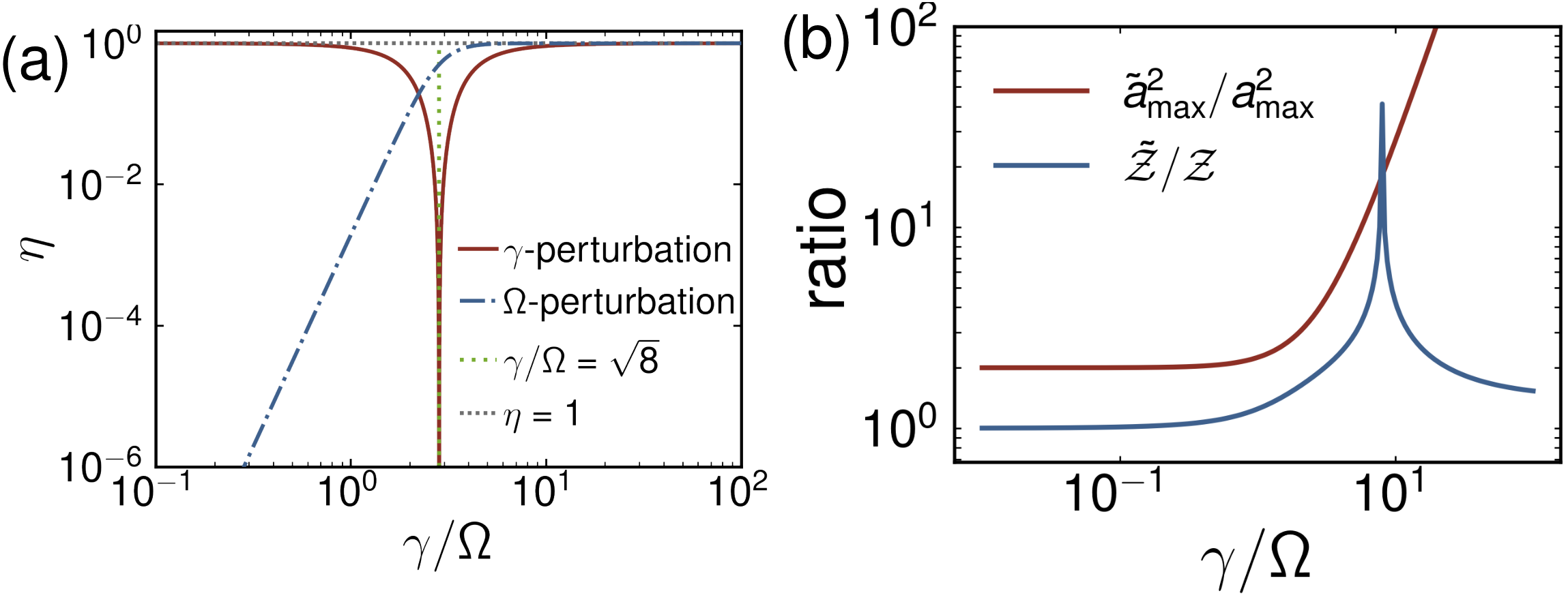}
 \caption{(a) Bound efficiency $\eta:= \mathrm{LHS}/(\mathsf{a}_{\max}^2\mathcal A+\mathcal Z)\le 1$ versus $\gamma/\Omega$ on a log--log scale for the two-level atom model with $\Delta=\bar{N}=0$. The red solid curve plots the case where $\gamma$ is perturbed and the blue dashdotted curve plots $\Omega$ being perturbed. The green dotted vertical line shows $\eta=0$ for $\gamma$-perturbation at $\gamma/\Omega=\sqrt{8}$ as expected by the analytical calculation. The gray dotted horizontal line represents the upper bound $\eta=1$ of the efficiency. (b) Operational overestimates in the same limiting case. Red: $\widetilde{\mathsf{a}_{\max}^2}/\mathsf{a}_{\max}^2$ for the $\gamma$-perturbation, with analytic form $(\gamma^2+8\Omega^2)/(4\Omega^2)$. This ratio tends to $2$ for $\gamma\ll\Omega$ and grows as ${\cal O}[(\gamma/\Omega)^2]$ for $\gamma\gg\Omega$. Blue: $\widetilde{\mathcal Z}/|\mathcal Z|$ for the $\Omega$-perturbation, where the exact $\mathcal Z=16/\gamma$ and the mixing-based $\widetilde{\mathcal Z}$ captures the ${\cal O}(1/\gamma)$ scaling.}
 \label{fig:limiting_and_operational_two_level}
\end{figure}

A limiting case in which every quantity can be solved analytically is obtained when the heat bath is at zero temperature $\beta\to\infty$ and the detuning $\Delta$ vanishes. The atom only undergoes spontaneous emission characterized by the jump operator $L_-$. We choose $\nu_{-}=1$ hence $\phi$ is the current rate. There are only two tunable parameters: $\gamma$ and $\Omega$. In the steady state, every quantity can be solved exactly \cite{landi2024},
\begin{align}
& \langle J\rangle=\frac{4\gamma\Omega^2}{\gamma^2+8\Omega^2},\quad \mathcal A=\langle J\rangle,\\
& \langle \langle J\rangle \rangle
=\frac{4\gamma\Omega^2(\gamma^4-8\gamma^2\Omega^2+64\Omega^4)}{(\gamma^2+8\Omega^2)^3}.
\end{align}
For a $\gamma$ perturbation,
\begin{align}
 \partial_\gamma\langle J\rangle
=\frac{4\Omega^2(-\gamma^2+8\Omega^2)}{(\gamma^2+8\Omega^2)^2},\  \mathsf a_{\max}^2=\frac{1}{\gamma^2},\ \mathcal Z=0.
\end{align}
so the bound reduces to the activity-controlled form,  and the bound efficiency is $\eta = 1- 1/(8\Omega^2/\gamma^2 + \gamma^2/8\Omega^2 -1)<1$.

 The efficiency vanishes at $\gamma/\Omega=\sqrt{8}$  where the current becomes first-order insensitive to the dissipation rate. Physically, increasing $\gamma$ enhances the emission rate per excitation but simultaneously suppresses the steady-state excited population; at $\gamma/\Omega=\sqrt{8}$ these two effects balance exactly.

For an $\Omega$ perturbation,
\begin{align}
 \partial_\Omega\langle J\rangle
=\frac{8\gamma^3\Omega}{(\gamma^2+8\Omega^2)^2},\  \mathsf a_{\max}^2=0, \ \mathcal Z=\frac{16}{\gamma},
\end{align}
so the bound efficiency becomes $\eta = 1 / (1 + 512 \Omega^6 / \gamma^6)$. Figure~\ref{fig:limiting_and_operational_two_level}(a) shows that the QR-KUR is always valid and can be extremely tight. The $\gamma$-perturbation isolates the activity contribution, whereas the $\Omega$-perturbation isolates the intertransition contribution, illustrating the equal importance of the two terms on the right-hand side of inequality~\eqref{eq:lindblad_rkur}.

\begin{figure*}[t]
 \centering
 \includegraphics[width=\textwidth]{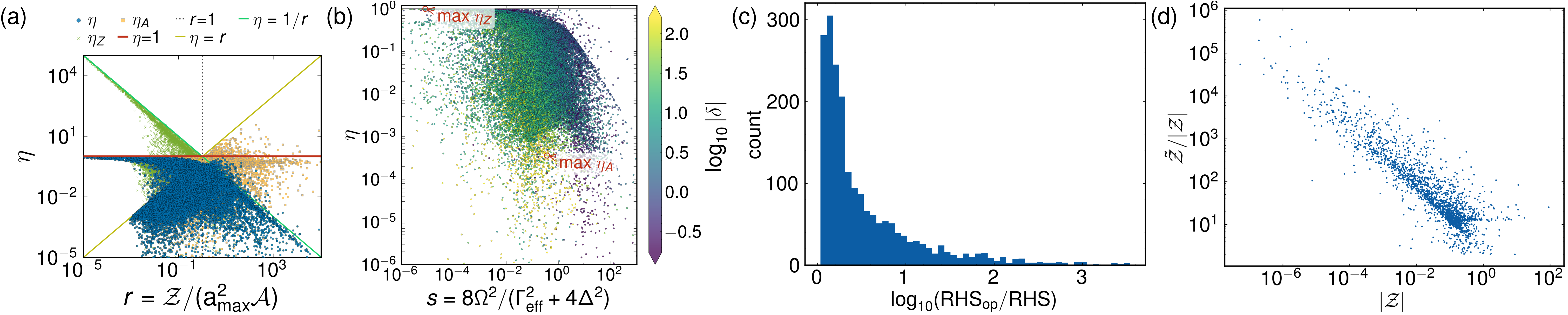}
 \caption{(a) Log-log plot versus the dominance ratio $r:=\mathcal Z/(\mathsf{a}_{\max}^2\mathcal A)$.
 In the sampled data shown here, $\mathcal Z>0$, so that both $r$ and
 $\eta_Z=\mathrm{LHS}/\mathcal Z$ are well-defined.
 Blue circles show the overall efficiency $\eta=\mathrm{LHS}/\mathrm{RHS}$, while yellow squares and green crosses show the single-term ratios
 $\eta_A=\mathrm{LHS}/(\mathsf{a}_{\max}^2\mathcal A)$ and $\eta_Z=\mathrm{LHS}/\mathcal Z$, respectively.
 The horizontal line marks $\eta=1$, and the vertical line marks $r=1$.
 The guide lines $y=r$ and $y=1/r$ indicate, respectively, the loci $\eta_A=r$ when $\eta_Z=1$ and $\eta_Z=1/r$ when $\eta_A=1$,
 making explicit that single-term failures ($\eta_A>1$ or $\eta_Z>1$) occur in opposite dominance regimes $r\gg 1$ and $r\ll 1$.
 (b) Log-log scatter of the overall efficiency $\eta$ versus the generalized saturation parameter
 $s=8\Omega^2/(\Gamma_{\rm eff}^2+4\Delta^2)$.
 Colors encode the detuning in linewidth units, $\log_{10}|\delta|$ with $\delta:=2\Delta/\Gamma_{\rm eff}$ and $\Gamma_{\rm eff}=\gamma(2\bar N+1)$.
 Circled points mark the samples that maximize the single-term ratios in the sample set.
 (c) Histogram of $\log_{10}(R_{\mathrm{RHS}})=\log_{10}(\mathrm{RHS}_{\rm op}/\mathrm{RHS})$ over the same $\omega$-perturbation ensemble, showing how much larger $\mathrm{RHS}_{\rm op}=\widetilde{\mathsf{a}_{\max}^2}\mathcal A+\widetilde{\mathcal Z}$ is than the exact RHS.
 (d) Scatter plot of $R_Z=\widetilde{\mathcal Z}/|\mathcal Z|$ versus the exact magnitude $|\mathcal Z|$ (log--log). The strongest overestimates occur when $|\mathcal Z|$ is small, consistent with the fact that $\mathcal Z$ can exhibit cancellations whereas the norm-based upper bound $\widetilde{\mathcal Z}$ discards such cancellations.}
 \label{fig:dominance_and_operational_scan}
\end{figure*}

\subsection{General parameter regime}

The general case allows both $\bar N$ and $\Delta$ to be nonzero.
Here the current is defined as the net current with $\nu_\pm=\pm 1$, unlike the emission-count observable used in the limiting case above where $\nu_-=1$.
The Lindbladian matrix can then be diagonalized numerically.
The effective linewidth is
$\Gamma_{\rm eff}=\gamma(2\bar N+1)$,
which provides the natural frequency scale against which both detuning and coherent drive should be compared.
A convenient dimensionless detuning is therefore the detuning in linewidth units, given by
$\delta=2\Delta/\Gamma_{\rm eff}$.
The coherent drive is characterized by the generalized saturation parameter
\begin{equation}
s=\frac{8\Omega^2}{\Gamma_{\rm eff}^2+4\Delta^2}.
\end{equation}
Equivalently, introducing the on-resonance saturation parameter
$s_0=8\Omega^2/\Gamma_{\rm eff}^2$, one has the identity $s=s_0/(1+\delta^2)$.
This representation makes explicit that a fixed value of $s$ does not correspond to a unique physical regime:
the same $s$ can arise from near-resonant driving with $|\delta|\ll 1$ and moderate $s_0$, or from far-detuned driving with $|\delta|\gg 1$ and a larger $s_0$.

The relative roles of the two contributions on the right-hand side of inequality~\eqref{eq:lindblad_rkur} are shown in Figs.~\ref{fig:dominance_and_operational_scan}(a) and (b).
In this ensemble $\mathcal Z>0$ throughout, so the dominance ratio
\begin{equation}
r := \frac{\mathcal Z}{\mathsf{a}_{\max}^2\mathcal A},
\end{equation}
as well as the overall efficiency and the two single-term ratios
\begin{align}
\eta := \frac{\mathrm{LHS}}{\mathsf{a}_{\max}^2\mathcal A+\mathcal Z},
\label{eq:eta_definitions_main}\ \eta_A := \frac{\mathrm{LHS}}{\mathsf{a}_{\max}^2\mathcal A},\ \eta_Z := \frac{\mathrm{LHS}}{\mathcal Z}.
\end{align}
Figure~\ref{fig:dominance_and_operational_scan}(a) shows that the combined ratio $\eta$ remains bounded by unity across the ensemble,
while the single-term ratios can exceed unity by orders of magnitude in opposite dominance regimes:
$\eta_A$ can be large when $r\gg 1$, where the $\mathcal Z$ term dominates the right-hand side. By contrast, $\eta_Z$ can be large when $r\ll 1$, where the $\mathsf{a}_{\max}^2\mathcal A$ term dominates.

Figure~\ref{fig:dominance_and_operational_scan}(b) presents the same ensemble in terms of the generalized saturation parameter $s$, with color encoding $\log_{10}|\delta|$ to distinguish near-resonant and far-detuned samples that may share the same $s$.
Near resonance, $|\delta|\ll 1$, coherent excitation is enhanced and the response is governed mainly by the competition between the Rabi drive $\Omega$ and the bath-induced linewidth $\Gamma_{\rm eff}$.
For $|\delta|\gg 1$, by contrast, the coherent response is suppressed by detuning, and achieving the same $s$ requires a larger on-resonance strength $s_0=8\Omega^2/\Gamma_{\rm eff}^2$.
This accounts for the systematic trend that smaller $|\delta|$ tends to produce larger $\eta$ and for the broader spread of efficiencies at fixed $s$ in the far-detuned regime.

The corner in which both $|\delta|$ and $s$ are small corresponds to weak, near-resonant driving, namely $s\ll 1$, so the stationary state remains close to equilibrium and the dynamics is governed mainly by the relaxation scale $\Gamma_{\rm eff}$.
In that regime the response entering the LHS and the noise and activity scales entering the RHS are controlled by the same dominant dissipative mode, and the inequality becomes nearly tight.
Increasing either $s$ or $|\delta|$ moves the system away from this simple near-resonant regime and lowers the observed efficiency.

The circled points mark the samples that maximize $\eta_A$ and $\eta_Z$ in the ensemble. For the present data set these occur at $\max \eta_A \approx 34.6$, with $r\approx 181$, $s\approx 0.98$, and $|\delta|\approx 0.35$, and at $\max \eta_Z \approx 7.17\times 10^{8}$, with $r\approx 1.39\times 10^{-9}$, $s\approx 8.03\times 10^{-6}$, and $|\delta|\approx 0.05$. They highlight that the most severe single-term failures occur in opposite RHS-dominance regions,
whereas the combined ratio $\eta$ remains uniformly bounded.

\begin{figure*}[t]
 \centering
 \includegraphics[width=\textwidth]{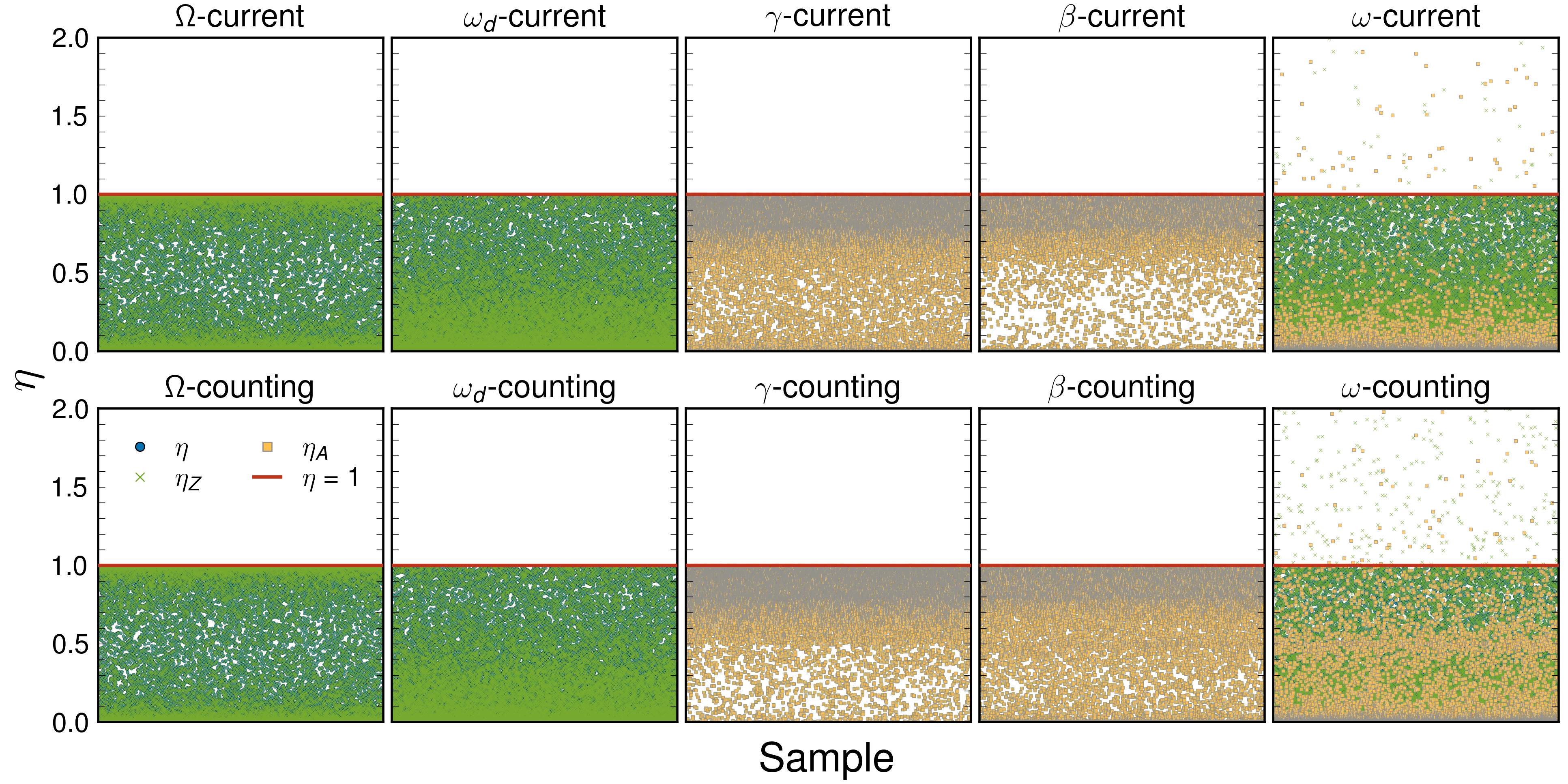}
 \caption{Bound efficiency $\eta$ in all perturbation and observable settings. In total, we have five parameters $\times$ two types of observables. Perturbation types: $\{\Omega, \omega_d\}$ only in Hamiltonian, $\eta=\eta_Z$; $\{\gamma, \beta\}$ only in jump operators, $\eta=\eta_A$; $\{\omega\}$ in both, $\eta$ is the combined ratio. Number of samples in each subfigure is 30000.}
 \label{fig:all_results}
\end{figure*}

A broader scan covers all five tunable parameters and both observable types.
Figure~\ref{fig:all_results} summarizes the resulting efficiencies for current and counting observables under $\Omega$, $\omega_d$, $\gamma$, $\beta$, and $\omega$ perturbations.
For all sampled settings the combined ratio $\eta$ remains below unity, while the single-term ratios may fail depending on which contribution dominates the right-hand side.
The validity of inequality~\eqref{eq:lindblad_rkur} is checked across all parameter-encoding classes considered here.

\section{Operational upper bounds}\label{sec:operational_bounds}
\subsection{General bounds from measured quantities}

The activity $\mathcal A=\sum_c\Tr{L_c(\theta)\pi(\theta)L_c^{\dagger}(\theta)}$ is directly accessible from the long-time quantum-jump record as the total click rate and, if desired, by channel \cite{DalibardCastinMolmer1992}.
In contrast, the prefactor $\mathsf a_{\max}^2$ is a quantity fixed by the model parameters that characterizes how sensitively the jump part of the dynamics depends on the parameter $\theta$.
In full generality,
\begin{align}
& \mathsf a_{\max}^2(\theta):=\max_c\mathsf a_c^2(\theta),
\label{eq:amax_general_operational}\\
& \mathsf a_c^2(\theta):=4\frac{\Tr{\partial_\theta L_c(\theta)\pi(\theta)\partial_\theta L_c^{\dagger}(\theta)}}{\Tr{L_c(\theta)\pi(\theta)L_c^{\dagger}(\theta)}}.
\end{align}
which depends only on how $\theta$ enters the jump operators in the chosen unraveling.
Accordingly, $\partial_\theta L_c(\theta)$ and hence $\mathsf a_{\max}^2$ is an input obtained from independent characterization of the device and its system--bath couplings, rather than a quantity to be inferred from the same trajectory used to estimate $\theta$ \cite{BraunsteinCaves1994,Paris2009QuantumEstimation}.
For the experimentally common multiplicative form $L_c(\theta)=g_c(\theta)L_c$ with $g_c(\theta)\in\mathbb{R}$, Eq.~\eqref{eq:amax_general_operational} reduces to the explicit expression $\mathsf a_{\max}^2=4\max_c(\partial_\theta\ln g_c)^2$.

An upper bound follows by combining channel-resolved click rates with bounds on the parameter dependence of the jump operators.
Let $\mathcal A_c(\theta):=\Tr{L_c(\theta)\pi(\theta)L_c^{\dagger}(\theta)}$ denote the steady-state click rate of channel $c$.
Then, for channels with nonzero click rate $\mathcal A_c(\theta)>0$,
\begin{align}
\tilde{\mathsf a}_c^2(\theta):=4\frac{\|\partial_\theta L_c(\theta)\|_{\rm op}^2}{\mathcal A_c(\theta)},
\label{eq:amax_relaxation}\  \tilde{\mathsf a}_{\max}^2(\theta):=\max_c\tilde{\mathsf a}_c^2(\theta).
\end{align}
provides an upper bound for Eq.~\eqref{eq:amax_general_operational}.
The derivation is given in Appendix~\ref{app:Z_upper_bound}.

The intertransition term $\mathcal Z$ is different in nature from $\mathcal A$.
Its exact expression is controlled by the Drazin pseudoinverse of the Liouvillian and hence by the relaxation spectrum on the decaying subspace.
For this reason, $\mathcal Z$ cannot, in general, be inferred from steady-state jump counts alone.
Evaluating $\mathcal Z$ therefore requires either a calibrated Markovian generator $\mathcal L(\theta)$ and its local derivatives entering $\mathcal K_{1,2}$, or time-resolved response measurements under a controlled modulation of $\theta$.
Experimentally demonstrated Lindblad-tomography and related generator-learning approaches provide routes to reconstruct effective Hamiltonian and jump operators from time-domain or tomographic data, enabling an a posteriori evaluation of $\mathcal Z$ within a validated Markovian description \cite{PhysRevApplied.18.064056,Onorati2023fittingquantumnoise}.

Let
$\hat{\mathcal Q}:=\mathbf 1-\kettbbra{\pi}{\mathbf 1}$
denote the projector onto the decaying subspace,
assuming $\bbrakett{\mathbf 1}{\pi}=1$ and a unique steady state.
One may further bound $\mathcal Z$ without explicitly constructing $\hat{\mathcal L}^{\mathrm D}$, under the standard stability assumption
\begin{equation}
\bigl\|e^{t\hat{\mathcal L}}\hat{\mathcal Q}\bigr\|_{2\to2}\le M e^{-\Delta t},\quad t\ge 0,
\label{eq:exp_mixing_assumption}
\end{equation}
where $\Delta$ is a mixing or relaxation scale on the decaying subspace and $M$ accounts for possible transient amplification when the $\hat{\mathcal Q}$-restricted generator is non-normal.
Under this assumption one obtains the upper bound
\begin{align}
|\mathcal Z|
&\le \widetilde{\mathcal Z}
:=\frac{8M}{\Delta}
\|\hat{\mathcal K}_1^\dagger\kett{1}\|_2
\|\hat{\mathcal Q}\hat{\mathcal K}_2\kett{\pi}\|_2.
\label{eq:Z_relaxation}
\end{align}
which trades spectral detail for a small number of a few inputs, notably the relaxation rate $\Delta$.
In the following, the two-level example provides a concrete test of how loose these upper bounds can become.

\subsection{Two-level illustration of the operational bounds}

In the analytically solvable two-level example with $\bar N=0$ and $\Delta=0$, for a $\gamma$-perturbation one has the exact $\mathsf{a}_{\max}^2=1/\gamma^2$ and $\mathcal Z=0$.
The relaxed contribution $\mathcal A\widetilde{\mathsf{a}_{\max}^2}$ reduces to $4\|\partial_\gamma L_-\|_{\rm op}^2=1/\gamma$ single channel,
while the exact $\mathsf{a}_{\max}^2\mathcal A$ is $4\Omega^2/(\gamma^2+8\Omega^2)$.
Thus the ratio of the relaxed to exact contribution is
\begin{equation}
\frac{\mathcal A\widetilde{\mathsf{a}_{\max}^2}}{\mathsf{a}_{\max}^2\mathcal A}
=\frac{\gamma^2+8\Omega^2}{4\Omega^2}.
\end{equation}
Equivalently, writing $p_e:=\Tr{\pi\sigma_+\sigma_-}=\mathcal A/\gamma$ for the steady-state excited-state population, this ratio is $1/p_e$.
Hence its weak-dissipation limit $2$ at $\gamma\ll\Omega$ reflects $p_e\to 1/2$ in the resonant driven steady state and does not indicate a missing numerical factor in the derivation.
Rather, it is the intrinsic overestimate produced by replacing the $\pi$-weighted matrix element by the operator norm in Eq.~\eqref{eq:amax_relaxation}.
In the strong-dissipation regime $\gamma\gg\Omega$, one finds
\begin{equation}
\frac{\mathcal A\widetilde{\mathsf{a}_{\max}^2}}{\mathsf{a}_{\max}^2\mathcal A}
=\frac{\gamma^2}{4\Omega^2}\left[1+{\cal O}\left((\Omega/\gamma)^2\right)\right],
\quad (\gamma\gg\Omega),
\end{equation}
showing explicitly that the relaxation \eqref{eq:amax_relaxation} can become parametrically loose even in a two-level system.

Inequality~\eqref{eq:Z_relaxation} replaces $|\mathcal Z|$ by a simple upper bound and is not uniformly tight.
Its natural scale is additive,
\begin{equation}
\widetilde{\mathcal Z} = {\cal O}\left(\frac{M}{\Delta}\|\hat{\mathcal K}\|^2\right),
\end{equation}
because $\mathcal Z$ is a time-integrated transient kernel controlled by relaxation on $\mathrm{Ran}(\hat{\mathcal Q})$.
A uniform relative tightness statement is impossible in general:
$\mathcal Z$ is a signed modal sum and may exhibit strong cancellations or even cross zero as parameters vary,
whereas inequality~\eqref{eq:Z_relaxation} is obtained by the estimate $|\mathcal Z|=2|\text{Re}  \Xi|\le 2|\Xi|$ together with Cauchy-Schwarz inequality and therefore discards such cancellations.
Consequently, $\widetilde{\mathcal Z}/|\mathcal Z|$ can become arbitrarily large near parameter values where $\mathcal Z\approx 0$, even though $\widetilde{\mathcal Z}$ remains at its natural scale ${\cal O}[(M/\Delta)\|\hat{\mathcal K}\|^2]$.

In the two-level example above, for an $\Omega$-perturbation one has $\mathsf{a}_{\max}^2=0$ and the exact $\mathcal Z=16/\gamma$.
In this case inequality~\eqref{eq:Z_relaxation} predicts the correct order ${\cal O}(1/\gamma)$ because the relaxation scale satisfies $\Delta={\cal O}(\gamma)$, and the constant prefactor can be checked numerically from the exact definitions of $\hat{\mathcal K}_{1,2}$ and $\hat{\mathcal Q}$.
This numerical check quantifies how loose the upper bound \eqref{eq:Z_relaxation} can be in the examples.

A second feature of Fig.~\ref{fig:limiting_and_operational_two_level}(b) is that the blue curve may develop a pronounced peak near $\gamma/\Omega\approx 8$.
This point should not be confused with the special value $\gamma/\Omega=\sqrt{8}$ in Fig.~\ref{fig:limiting_and_operational_two_level}(a).
In the resonant limit $\bar N=0$ and $\Delta=0$, the nonzero eigenvalues of the Bloch-linearized generator are
$\lambda_1=-\gamma/2$ and $\lambda_{\pm}=-3\gamma/4\pm \sqrt{\gamma^2/16-4\Omega^2}$.
Thus $\gamma/\Omega=8$ marks the underdamped--overdamped crossover, where the pair $\lambda_{\pm}$ changes from complex to real.
Near this crossover the non-normal prefactor $M$ entering inequality~\eqref{eq:Z_relaxation} can become enhanced, which enlarges $\widetilde{\mathcal Z}/|\mathcal Z|$ even though the exact $\mathcal Z=16/\gamma$ remains regular.
By contrast, $\gamma/\Omega=\sqrt{8}$ is the point where $\partial_{\gamma}\langle J\rangle=0$, namely the extremum of the steady-state emission rate.
The two values therefore reflect different mechanisms: a relaxation-spectrum crossover for the peak in the blue curve and a response extremum for the zero of the red curve in Fig.~\ref{fig:limiting_and_operational_two_level}(a).

A broader random scan shows the same pattern beyond the resonant limiting case.
For the $\omega$-perturbation ensemble underlying Figs.~\ref{fig:dominance_and_operational_scan}(a) and (b), we evaluate both the exact and relaxed right-hand sides and record
\begin{equation}
R_{\mathrm{RHS}}
:=
\frac{\widetilde{\mathsf{a}_{\max}^2}\mathcal A+\widetilde{\mathcal Z}}{\mathsf{a}_{\max}^2\mathcal A+\mathcal Z},
\quad
R_Z:=\frac{\widetilde{\mathcal Z}}{|\mathcal Z|}.
\end{equation}
Figures~\ref{fig:dominance_and_operational_scan}(c) and (d) show that $R_{\mathrm{RHS}}$ is typically moderate, often within a few orders of magnitude, whereas $R_Z$ can become very large when $|\mathcal Z|$ is small.
This is expected because $\mathcal Z$ is a signed modal sum that can exhibit strong cancellations or approach zero, while $\widetilde{\mathcal Z}$ follows from the estimate $|\mathcal Z|=2|\text{Re} \Xi|\le 2|\Xi|$ and therefore discards those cancellations.
The operational replacement is therefore useful as a coarse-grained accessible substitute for the exact RHS, but it should not be interpreted as a uniformly tight estimator of $\mathcal Z$ itself.

\section{Symmetry-resolved structure of the intertransition term}\label{sec:group_theory}

While the operational bound above captures the accessible scale of the exact right-hand side, it does not reveal which decaying sectors are responsible for the intertransition term itself.
This motivates a structural analysis of $\mathcal Z$, to which we turn next from the viewpoint of symmetry.
In particular, once $\mathcal Z$ is identified as arising from perturbation-induced couplings between the steady-state sector and decaying sectors, a natural question is which relaxation channels can actually contribute and how such contributions are constrained by symmetry.
To answer this question, we now develop a representation-theoretic decomposition of $\mathcal Z$, which resolves the intertransition term into symmetry sectors and yields selection rules for the allowed contributions. A detailed derivation is given in Appendix~\ref{app:group_decomposition}. In particular, Eq.~\eqref{eq:Z_definition} is the exact real form of the intertransition term under the GKSL assumptions.

Let $G$ be a compact group with a unitary representation $g\mapsto \hat U_g$ on Liouville space such that
\begin{equation}
[\hat{\mathcal L},\hat U_g]=0,
\quad \forall g\in G.
\label{eq:L_commute_Ug_group}
\end{equation}
Then Liouville space has an isotypic decomposition
\begin{align}
 \mathcal H_{\rm L}=\bigoplus_{\lambda\in\widehat G}\mathcal H_\lambda,
\label{eq:isotypic_decomp_main}\  \mathcal H_\lambda\cong V_\lambda\otimes M_\lambda.
\end{align}
where $V_\lambda$ is the carrier space of the irrep $\lambda$ and $M_\lambda$ is its multiplicity space.
Denote by $\hat P_\lambda$ the corresponding projectors.
Since both $\hat{\mathcal L}$ and $\hat{\mathcal L}^{\mathrm D}$ commute with the group action, they preserve every isotypic block.
Moreover, on each block one has the reduced-resolvent form
\begin{equation}
\hat P_\lambda \hat{\mathcal L}^{\mathrm D}\hat P_\lambda
=
\mathbf 1_{V_\lambda}\otimes R_\lambda,
\label{eq:LD_reduced_resolvent_main}
\end{equation}
with $R_\lambda$ acting only on the multiplicity space $M_\lambda$.
If the steady state is unique, then both $\kett{\pi}$ and $\bbra{\mathbf 1}$ belong to the trivial isotypic component, denote this component by $\lambda=0$.
Inserting the isotypic resolution of identity into the complex bilinear form inside Eq.~\eqref{eq:Z_definition} yields the exact decomposition
\begin{align}
&\mathcal Z
=\sum_{\lambda\in\widehat G}\mathcal Z^{(\lambda)},
\label{eq:Z_isotypic_exact_main}\\
&\mathcal Z^{(\lambda)}
:=-8{\rm Re}\left[
\bbra{\mathbf 1}
\hat{\mathcal K}_1
\hat P_\lambda
\hat{\mathcal L}^{\mathrm D}
\hat P_\lambda
\hat{\mathcal K}_2
\kett{\pi}
\right].
\end{align}
Thus $\mathcal Z$ is decomposed exactly into symmetry-resolved sector contributions, and all dynamical information within an allowed sector is compressed into the reduced resolvent $R_\lambda$ on the multiplicity space.

This immediately yields a complete classification of symmetry-allowed sectors.
Define the ket-side and bra-side symmetry supports
\begin{align}
    &S_{\rm R}(\hat{\mathcal K}_2)
    :=
    \{\lambda\in\widehat G:\ \hat P_\lambda \hat{\mathcal K}_2\kett{\pi}\neq 0\},
    \nonumber\\
    &S_{\rm L}(\hat{\mathcal K}_1)
    :=
    \{\lambda\in\widehat G:\ \bbra{\mathbf 1}\hat{\mathcal K}_1\hat P_\lambda\neq 0\}.
    \label{eq:support_sets_main}
\end{align}
Then $\mathcal Z^{(\lambda)}=0$ for every $\lambda\notin S_{\rm L}(\hat{\mathcal K}_1)\cap S_{\rm R}(\hat{\mathcal K}_2)$.
Equivalently, only sectors contained in this intersection are symmetry-allowed contributors to $\mathcal Z$.

The commuting case considered previously is recovered immediately.
If
\begin{equation}
[\hat{\mathcal K}_i,\hat U_g]=0,
\quad \forall g\in G,
\quad i\in\{1,2\},
\label{eq:K_commute_Ug_group}
\end{equation}
then each $\hat{\mathcal K}_i$ preserves every isotypic block, so $S_{\rm R}(\hat{\mathcal K}_2)=S_{\rm L}(\hat{\mathcal K}_1)=\{0\}$.
Hence only the trivial isotypic component can contribute to $\mathcal Z$, and every nontrivial irrep is excluded exactly.
The previous sector-selection rule is therefore the trivial-irrep corollary of Eq.~\eqref{eq:Z_isotypic_exact_main}.

A different situation arises when the perturbation does not commute with the symmetry but transforms in irreducible symmetry channels.
Suppose that $\hat{\mathcal K}_2$ contains an irreducible multiplet $\{\hat S^{(\mu_2)}_b\}_{b=1}^{d_{\mu_2}}$ transforming as
\begin{equation}
\hat U_g \hat S^{(\mu_2)}_b \hat U_g^{-1}
=
\sum_{c=1}^{d_{\mu_2}} D^{(\mu_2)}_{cb}(g)\hat S^{(\mu_2)}_c,
\label{eq:covariant_channel_main_K2}
\end{equation}
and similarly that $\hat{\mathcal K}_1$ contains an irreducible multiplet $\{\hat T^{(\mu_1)}_a\}_{a=1}^{d_{\mu_1}}$.
Because $\kett{\pi}$ is invariant, the image $\hat S^{(\mu_2)}_b\kett{\pi}$ lies entirely in the $\mu_2$-isotypic component.
On the bra side, $\bbra{\mathbf 1}\hat T^{(\mu_1)}_a$ can couple only to the contragredient channel $\mu_1^*$.
Therefore the pair $(\mu_1,\mu_2)$ contributes only if
\begin{equation}
\mu_2\cong \mu_1^*
\quad\Longleftrightarrow\quad
\mathbf 1\subset \mu_1\otimes \mu_2.
\label{eq:group_selection_rule_main}
\end{equation}
This is the symmetry-channel selection rule for the intertransition term.
For reducible perturbations one simply decomposes $\hat{\mathcal K}_1$ and $\hat{\mathcal K}_2$ into irreducible channels and takes the union of all allowed contragredient pairs.

Equation~\eqref{eq:LD_reduced_resolvent_main} also shows where the sign of an allowed sector contribution is decided.
Choose an orthonormal basis on $\mathcal H_\lambda\cong V_\lambda\otimes M_\lambda$ and expand the ket-side and bra-side couplings into the $\lambda$ block as detailed in Appendix~\ref{app:group_decomposition}.
Then one can write
\begin{equation}
\mathcal Z^{(\lambda)}
=
-8{\rm Re}\left[\sum_{m=1}^{d_\lambda}\sum_{\beta,\beta'=1}^{\dim M_\lambda}
 d^*_{m\beta}(R_\lambda)_{\beta\beta'}c_{m\beta'}\right].
\label{eq:negative_Z_block_formula_main}
\end{equation}
Thus symmetry determines which blocks are allowed, whereas the sign inside an allowed block is controlled by the phase-sensitive reduced pairing between the left and right perturbative couplings through the reduced resolvent $R_\lambda$.
A negative contribution therefore means that the perturbation-induced excursion into a decaying quantum sector returns with a phase relation that enhances the response more than it increases the activity-only cost.
In this sense $\mathcal Z<0$ represents a quantum enhancement beyond the activity term alone.

A particularly transparent corollary is obtained when only one symmetry-allowed isotypic component $\lambda$ contributes and, within this block, only one complex-conjugate pair of decaying modes $\lambda_\pm=-\alpha\pm i\beta$ with $\alpha>0$ enters the reduced pairing.
Writing the corresponding modal coefficient as $c_+=|c|e^{i\phi}$, so that the partner coefficient is $c_- = c_+^*$, one finds
\begin{equation}
\mathcal Z^{(\lambda)}
=
-16{\rm Re}\left(\frac{c_+}{\lambda_+}\right)
=
-\frac{16|c|}{\alpha^2+\beta^2}\bigl(-\alpha\cos\phi+\beta\sin\phi\bigr).
\label{eq:negative_Z_single_pair_main}
\end{equation}
Hence $\mathcal Z^{(\lambda)}<0$ if and only if ${\rm Re}(c_+/\lambda_+)>0$, equivalently if and only if $-\alpha\cos\phi+\beta\sin\phi>0$.
This is the exact negative-$\mathcal Z$ criterion in the single-pair regime. The explicit qubit counterexample in Appendix~\ref{app:signofZ} realizes this mechanism through a single contributing parity-odd block of the $\mathbb Z_2$ decomposition generated by conjugation with $\sigma_x$.

For an Abelian $U(1)$ symmetry the irreps are one-dimensional charge sectors, and Eq.~\eqref{eq:group_selection_rule_main} reduces to a charge-selection rule.
Neutral perturbations ($q=0$) can probe only neutral decaying sectors, whereas a perturbation carrying charge $q$ can contribute only through decaying sectors of charge $q$ on the ket side and charge $-q$ on the bra side.
The driven two-level model provides a simple realization of this structure: dissipative perturbations remain in the neutral sector, whereas a transverse Hamiltonian perturbation activates the coherence sectors of opposite charge.

\section{Discussion}\label{sec:discussion}

The QR-KUR proved in the main text is formulated for a unique steady state $\pi$, equivalently for a one-dimensional kernel of $\hat{\mathcal L}$.
Multiple steady states may arise from strong symmetries or conserved quantities, and then the long-time state can depend on the initial condition \cite{BucaProsen2012,AlbertJiang2014,Latune2019}.
The Cram\'er--Rao step for the integrated observable $\Phi(T)$ remains valid,
\begin{align}
\frac{(\partial_\theta\langle\Phi(T)\rangle)^2}{\langle\langle\Phi(T)\rangle\rangle}
&\le F(\theta;T).
\label{eq:CR_general_nonunique}
\end{align}
but the reduction to a universal rate bound additionally requires linear-in-$T$ scaling of both the mean and the variance.
If the long-time dynamics decomposes into invariant components $\alpha$ with probabilities $p_\alpha$ and conditional scalings
\begin{equation}
\label{eq:time_extensivity_assumption}
\mathbb E[\Phi(T)|\alpha]=T\mu_\alpha+o(T),\ 
\mathrm{Var}[\Phi(T)|\alpha]=T\sigma_\alpha^2+o(T).
\end{equation}
then the law of total variance yields
\begin{align}
\mathrm{Var}[\Phi(T)]
&=T^2\Bigl(\sum_\alpha p_\alpha\mu_\alpha^2-\bigl(\sum_\alpha p_\alpha\mu_\alpha\bigr)^2\Bigr)\nonumber\\
&\quad +T\sum_\alpha p_\alpha\sigma_\alpha^2+o(T^2).
\label{eq:variance_mixture_nonunique}
\end{align}
Whenever the component-dependent rates $\mu_\alpha$ are not all identical, the leading fluctuations are of order $T^2$ rather than $T$.
In that situation a universal steady-state rate QR-KUR written in terms of a single stationary state is no longer natural.

If the preparation and the dynamics restrict the evolution to a single invariant sector and the steady state is unique within that sector, then the same derivation applies sector-wise.
If, by contrast, the preparation allows a mixture of invariant steady components with different steady rates, then one should either formulate the bound conditionally on the sector label or work directly with the integrated observable $\Phi(T)$ through inequality~\eqref{eq:CR_general_nonunique}.

For the steady state of a Lindblad equation, the response precision of a measured observable is bounded by quantum transitions tied to the steady-state sector.
The quantum jumps within the steady-state sector contribute to the conventional quantum dynamical activity, whereas perturbation-induced couplings between the steady-state sector and the decaying sectors generate a genuinely quantum contribution that is absent in the classical limit.
Simple situations in which either of the two contributions vanishes are identified, and the bound is verified in a driven two-level atom with dissipation.

The bound is formulated for a unique steady state.
The group-theoretic analysis identifies symmetry channels that can contribute to $\mathcal Z$ in that setting, while the nonunique steady-state issue discussed here concerns the boundary of the same framework.
Extensions to strong symmetries and to dynamics with conserved quantities or multiple zero modes remain of interest, since the asymptotic state can then depend on the initial condition \cite{AlbertJiang2014,BucaProsen2012}.

\begin{acknowledgments}
We acknowledge Zongping Gong, Masaya Nakagawa, Haoyu Guan, Xiuhao Deng, and Tingfei Li for fruitful discussions. 
This work was supported by the Scientific Research Start-up Foundation of Xihua University under Grant No. Z241064.
\end{acknowledgments}

\section*{DATA AVAILABILITY}

The code and the data used to plot the figures are available from the authors upon reasonable requests.

\appendix

\section{Proof of the QR-KUR}\label{app:proof_qrkur}

This appendix recalls only the ingredients from Ref.~\cite{gammelmark2014} that are needed in the steady-state monitored GKSL setting and then derives inequality~\eqref{eq:lindblad_rkur} in the notation of the main text.

\subsection{Vectorization and long-time QFI}

Liouville-space notation is used throughout the appendices. With stack-column vectorization,
\begin{equation}
\mathrm{vec}(ABC)=(C^{\mathbb{T}}\otimes A)\mathrm{vec}(B).
\end{equation}
so that
\begin{align}
    & \mathrm{vec}(L\rho L^{\dagger})=(L^*\otimes L)\kett{\rho},
    \\ &
    \mathrm{vec}([H,\rho])=(\mathbf 1\otimes H-H^{\mathbb T}\otimes \mathbf 1)\kett{\rho}.
\end{align}
The matrix representation of the Lindbladian \eqref{eq:Lindbladian} is therefore
\begin{align}
 \hat{\mathcal L}
 ={}&-i(\mathbf 1\otimes H-H^{\mathbb T}\otimes \mathbf 1)\nonumber\\
 &+\sum_c\left[L_c^*\otimes L_c-\frac12\mathbf 1\otimes L_c^{\dagger}L_c-\frac12(L_c^{\dagger}L_c)^{\mathbb T}\otimes \mathbf 1\right].
\end{align}
Assume a diagonalizable finite-dimensional GKSL generator with a unique steady state $\pi$, a simple zero mode, and $\text{Re} \lambda_j<0$ for all nonzero eigenvalues. With
\begin{equation}
\hat{\mathcal L}\kett{x_j}=\lambda_j\kett{x_j},
\quad
\bbra{y_j}\hat{\mathcal L}=\lambda_j\bbra{y_j}.
\end{equation}
we choose $\kett{x_0}=\kett{\pi}$ and $\bbra{y_0}=\bbra{\mathbf 1}$. The biorthogonal decomposition then yields
\begin{equation}
\hat{\mathcal L}^{\mathrm D}:=\sum_{j\ne 0}\frac{1}{\lambda_j}\kettbbra{x_j}{y_j}.
\end{equation}
which is the Drazin pseudoinverse on the decaying subspace \cite{Drazin1958}.

For two parameter values $\theta_1$ and $\theta_2$, introduce the generalized Lindbladian
\begin{align}
 {\mathcal L}(\theta_1,\theta_2)\rho 
  :={}&-iH(\theta_1)\rho+i\rho H(\theta_2)+\sum_c L_c(\theta_1)\rho L_c(\theta_2)^\dagger\nonumber\\
 &-\frac12\sum_c\left[L_c(\theta_1)^\dagger L_c(\theta_1)\rho+\rho L_c(\theta_2)^\dagger L_c(\theta_2)\right].
\end{align}
In the long-time limit, Ref.~\cite{gammelmark2014} gives the QFI of the monitored trajectory as
\begin{equation}
\label{eq:FIeig}
\mathcal F(\theta)=\left.4T\partial_{\theta_1}\partial_{\theta_2}\text{Re} \lambda(\theta_1,\theta_2)\right|_{\theta_1=\theta_2=\theta}.
\end{equation}
where $\lambda(\theta_1,\theta_2)$ is the eigenvalue of $\mathcal L(\theta_1,\theta_2)$ that is smoothly connected to the stationary eigenvalue at $(\theta,\theta)$.
Applying the perturbative eigenvalue formula of Ref.~\cite{gammelmark2014} to this generalized Lindbladian gives
\begin{equation}
4\left.\partial_{\theta_1}\partial_{\theta_2}\lambda(\theta_1,\theta_2)\right|_{\theta_1=\theta_2=\theta}
=\mathcal X+\Xi+\Xi^*.
\end{equation}
with
\begin{equation}
\mathcal X:=4\sum_c\Tr{\partial_\theta L_c \pi \partial_\theta L_c^\dagger}\in\mathbb R_0^+.
\end{equation}
and
\begin{align}
 \Xi&:=-4\bbra{\mathbf 1}\hat{\mathcal K}_1\hat{\mathcal L}^{\mathrm D}\hat{\mathcal K}_2\kett{\pi}
 \nonumber\\&=-4\sum_{j\ne 0}\frac{1}{\lambda_j}\bbra{\mathbf 1}\hat{\mathcal K}_1\kettbbra{x_j}{y_j}\hat{\mathcal K}_2\kett{\pi}.
\label{eq:Xi_spectral_methods}
\end{align}
The perturbation superoperators are
\begin{align}
 \mathcal K_1\rho&:=\left.\partial_{\theta_1}\mathcal L(\theta_1,\theta_2)\rho\right|_{\theta_1=\theta_2=\theta}\nonumber\\&
=-i(\partial_\theta H_{\mathrm eff})\rho+\sum_c(\partial_\theta L_c)\rho L_c^\dagger,
\label{eq:K1_appendix}
\\
 \mathcal K_2\rho&:=\left.\partial_{\theta_2}\mathcal L(\theta_1,\theta_2)\rho\right|_{\theta_1=\theta_2=\theta}\nonumber\\&
=+i\rho(\partial_\theta H_{\mathrm eff})^\dagger+\sum_c L_c\rho \partial_\theta L_c^\dagger.
\label{eq:K2_appendix}
\end{align}
where $H_{\mathrm eff}=H-\frac{i}{2}\sum_cL_c^\dagger L_c$.
Defining
\begin{equation}
\mathcal Z:=\Xi+\Xi^* =2\text{Re} \Xi,
\end{equation}
the long-time QFI becomes
\begin{equation}
\label{eq:qfi_continuous_measurement}
\mathcal F(\theta)=T(\mathcal X+\mathcal Z).
\end{equation}

\subsection{Derivation of the bound}

The quantum Cram\'er--Rao bound gives
\begin{equation}
\label{eq:QCRB}
\langle \langle\Phi\rangle \rangle_{\theta}\ge \frac{(\partial_\theta\langle\Phi\rangle_{\theta})^2}{\mathcal F(\theta)}.
\end{equation}
In the steady state,
\begin{equation}
\partial_\theta\langle\Phi(T)\rangle=T\partial_\theta\langle\phi\rangle,
\quad
\langle \langle\Phi(T)\rangle \rangle=T\langle \langle\phi\rangle \rangle.
\end{equation}
Combining these relations with Eq.~\eqref{eq:qfi_continuous_measurement} yields
\begin{equation}
\frac{(\partial_\theta\langle\phi\rangle)^2}{\langle \langle\phi\rangle \rangle}
\le \mathcal X+\mathcal Z.
\end{equation}
It remains to bound $\mathcal X$. By the definition of $\mathsf a_{\max}^2$ in the main text,
\begin{equation}
4\Tr{\partial_\theta L_c \pi \partial_\theta L_c^\dagger}
\le \mathsf a_{\max}^2\Tr{L_c\pi L_c^\dagger}.
\end{equation}
for every channel included in the maximum. Summing over $c$ gives
\begin{align}
 \mathcal X
&=4\sum_c\Tr{\partial_\theta L_c \pi \partial_\theta L_c^\dagger}
\nonumber\\& \le \mathsf a_{\max}^2\sum_c\Tr{L_c\pi L_c^\dagger}
=\mathsf a_{\max}^2\mathcal A.
\end{align}
Substitution into the previous inequality yields Eq.~\eqref{eq:lindblad_rkur}.

\section{Proofs of auxiliary properties of the intertransition term}\label{app:Z_properties}

\subsection{Sign of $\mathcal Z$}
\label{app:signofZ}
A simple GKSL qubit example also shows that $\mathcal Z$ need not be non-negative. Consider
\begin{align}
 H=\sigma_x,
\label{eq:counterexample_model_methods}\ L_1=\sigma_-,\ L_2=\sigma_+,\ \pi=\frac{\mathbf 1}{2}.
\end{align}
and perturb only the jump operators according to
\begin{equation}
\label{eq:counterexample_perturbation_methods}
\partial_\theta H=0,\quad
\partial_\theta L_1=\partial_\theta L_2:=-(\sigma_x+\sigma_z).
\end{equation}
Then $\partial_\theta H_{\mathrm eff}=i\mathbf 1$, so that
\begin{equation}
\label{eq:counterexample_K12_methods}
\mathcal K_1(\rho)=\rho+(\partial_\theta L)\rho\sigma_x,
\quad
\mathcal K_2(\rho)=\rho+\sigma_x\rho(\partial_\theta L).
\end{equation}
Evaluated at $\pi=\mathbf 1/2$, this yields
\begin{equation}
\label{eq:counterexample_Kpi_methods}
\mathcal K_1(\pi)=-\frac{i}{2}\sigma_y,
\quad
\mathcal K_2(\pi)=+\frac{i}{2}\sigma_y.
\end{equation}
On the invariant subspace $\mathrm{span}\{\sigma_y,\sigma_z\}$ one has
\begin{equation}
\label{eq:counterexample_Lyz_methods}
\mathcal L(\sigma_y)=2\sigma_z-\sigma_y,
\quad
\mathcal L(\sigma_z)=-2\sigma_y-2\sigma_z.
\end{equation}
Solving $\mathcal L(X)=\sigma_y$ within this subspace gives
\begin{equation}
\label{eq:counterexample_Drazin_methods}
\mathcal L^{\mathrm D}(\sigma_y)=-\frac{1}{3}(\sigma_y+\sigma_z),
\quad
\mathcal L^{\mathrm D}(\mathcal K_2(\pi))=-\frac{i}{6}(\sigma_y+\sigma_z).
\end{equation}
Substituting into Eq.~\eqref{eq:Xi_spectral_methods} then gives
\begin{equation}
\label{eq:counterexample_Z_negative}
\Xi=-\frac{4}{3},
\quad
\mathcal Z=2\text{Re} (\Xi)=-\frac{8}{3}<0.
\end{equation}
Thus $\mathcal Z$ is not sign definite in general. This example also realizes the single-pair negative-$\mathcal Z$ criterion discussed in Appendix~\ref{app:group_decomposition}.

\subsection{Operational upper bound}\label{app:Z_upper_bound}

The single-term operational relaxation \eqref{eq:Z_relaxation} follows from the time-integral representation of the Drazin pseudoinverse. Starting from
\begin{equation}
\Xi:=-4\bbra{\mathbf 1}\hat{\mathcal K}_1\hat{\mathcal L}^{\mathrm D}\hat{\mathcal K}_2\kett{\pi},
\quad
\mathcal Z=2\text{Re} \Xi.
\end{equation}
and using
\begin{equation}
\hat{\mathcal L}^{\mathrm D}\hat{\mathcal Q}
=-\int_0^\infty dt\ e^{t\hat{\mathcal L}}\hat{\mathcal Q},
\end{equation}
one obtains
\begin{align}
& \Xi=4\int_0^\infty dt\bbra{u_1}e^{t\hat{\mathcal L}}\kett{v_2},\\
& \bbra{u_1}:=\bbra{\mathbf 1}\hat{\mathcal K}_1,\  \kett{v_2}:=\hat{\mathcal Q}\hat{\mathcal K}_2\kett{\pi}.
\end{align}
Cauchy--Schwarz in Hilbert--Schmidt space together with the exponential-mixing assumption \eqref{eq:exp_mixing_assumption} gives
\begin{align}
\bigl|\bbra{u_1}e^{t\hat{\mathcal L}}\kett{v_2}\bigr|
&\le M e^{-\Delta t}
\|\hat{\mathcal K}_1^\dagger\kett{1}\|_2
\|\hat{\mathcal Q}\hat{\mathcal K}_2\kett{\pi}\|_2.
\end{align}
Integrating over $t\ge0$ and using $|\mathcal Z|\le 2|\Xi|$ yields
\begin{equation}
|\mathcal Z|
\le
\widetilde{\mathcal Z}
:=
\frac{8M}{\Delta}
\|\hat{\mathcal K}_1^\dagger\kett{1}\|_2
\|\hat{\mathcal Q}\hat{\mathcal K}_2\kett{\pi}\|_2,
\label{eq:Z_relaxation_methods}
\end{equation}
which is inequality~\eqref{eq:Z_relaxation}.

For any density matrix $\pi\ge 0$ with $\Tr{\pi}=1$,
\begin{align}
    \Tr{\partial_\theta L_c \pi \partial_\theta L_c^\dagger}
    &=
    \Tr{\pi^{1/2}(\partial_\theta L_c^\dagger \partial_\theta L_c)\pi^{1/2}}
    \nonumber\\ & \le
    \|\partial_\theta L_c\|_{\rm op}^2.
\end{align}
Hence
\begin{align}
    \mathsf a_c^2(\theta)
    & =
    4\frac{\Tr{\partial_\theta L_c(\theta)\pi(\theta)\partial_\theta L_c^\dagger(\theta)}}
    {\Tr{L_c(\theta)\pi(\theta)L_c^\dagger(\theta)}}
    \nonumber\\ & 
    \le
    4\frac{\|\partial_\theta L_c(\theta)\|_{\rm op}^2}{\mathcal A_c(\theta)}
    =
    \tilde{\mathsf a}_c^2(\theta).
\end{align}

\section{Group-theoretic decomposition and symmetry-channel classification of the intertransition term}\label{app:group_decomposition}

The representation-theoretic form used in the main text is derived below.
Under the GKSL assumptions the intertransition term has the exact real representation
\begin{equation}
\mathcal Z
=
-8{\rm Re}\left[\bbra{\mathbf 1}\hat{\mathcal K}_1\hat{\mathcal L}^{\mathrm D}\hat{\mathcal K}_2\kett{\pi}\right].
\label{eq:Z_from_Z1_methods}
\end{equation}
The complex bilinear form inside the real part is analyzed first, and $\mathcal Z$ is recovered at the end.

Let $G$ be a compact group with a unitary representation $g\mapsto \hat U_g$ on Liouville space such that
\begin{equation}
[\hat{\mathcal L},\hat U_g]=0,
\quad \forall g\in G.
\label{eq:L_commute_Ug_methods}
\end{equation}
Because finite-dimensional unitary representations of compact groups are completely reducible, Liouville space decomposes as
\begin{equation}
\mathcal H_{\rm L}
=
\bigoplus_{\lambda\in\widehat G}\mathcal H_\lambda,
\quad
\mathcal H_\lambda\cong V_\lambda\otimes M_\lambda,
\label{eq:isotypic_decomp_methods}
\end{equation}
where $V_\lambda$ is the carrier space of the irrep $\lambda$ and $M_\lambda$ is its multiplicity space.
Let $\hat P_\lambda$ be the projector onto $\mathcal H_\lambda$.
Since $\hat{\mathcal L}$ commutes with every $\hat U_g$, each isotypic component is invariant and therefore
\begin{equation}
\hat P_\nu \hat{\mathcal L}\hat P_\lambda
=
\delta_{\nu\lambda}\hat P_\lambda \hat{\mathcal L}.
\label{eq:L_block_isotypic_methods}
\end{equation}
The same holds for $\hat{\mathcal L}^{\mathrm D}$.
Indeed, $\hat{\mathcal L}^{\mathrm D}$ is a spectral function of $\hat{\mathcal L}$ on the decaying subspace, so it commutes with the same group action, and hence
\begin{equation}
\hat P_\nu \hat{\mathcal L}^{\mathrm D}\hat P_\lambda
=
\delta_{\nu\lambda}\hat P_\lambda \hat{\mathcal L}^{\mathrm D}.
\label{eq:LD_block_isotypic_methods}
\end{equation}

Equation~\eqref{eq:LD_block_isotypic_methods} can be refined on each isotypic component.
Fix $\lambda$.
On $\mathcal H_\lambda\cong V_\lambda\otimes M_\lambda$, the group acts as $D^{(\lambda)}(g)\otimes \mathbf 1_{M_\lambda}$.
The restricted operator
\begin{equation}
\hat X_\lambda:=\hat P_\lambda \hat{\mathcal L}^{\mathrm D}\hat P_\lambda
\label{eq:Xlambda_def_methods}
\end{equation}
is an intertwiner of this representation because $\hat X_\lambda$ commutes with every $\hat U_g$.
By the standard commutant structure of an isotypic component, equivalently by Schur's lemma applied to the irrep factor, there exists a unique operator $R_\lambda$ on $M_\lambda$ such that
\begin{equation}
\hat P_\lambda \hat{\mathcal L}^{\mathrm D}\hat P_\lambda
=
\mathbf 1_{V_\lambda}\otimes R_\lambda.
\label{eq:reduced_resolvent_methods}
\end{equation}
Thus the full Drazin pseudoinverse is reduced, on each isotypic block, to an operator acting only on the multiplicity space.

If the steady state is unique, then $\kett{\pi}$ is symmetry-invariant.
Indeed, $\hat U_g\kett{\pi}$ is also a steady state because
$\hat{\mathcal L}\hat U_g\kett{\pi}
=
\hat U_g\hat{\mathcal L}\kett{\pi}=0$,
and uniqueness implies $\hat U_g\kett{\pi}=\kett{\pi}$.
Also, $\bbra{\mathbf 1}\hat U_g=\bbra{\mathbf 1}$ because $\hat U_g$ is trace-preserving.
Hence
\begin{equation}
\hat P_0\kett{\pi}=\kett{\pi},
\quad
\bbra{\mathbf 1}\hat P_0=\bbra{\mathbf 1},
\label{eq:steady_trivial_isotypic_methods}
\end{equation}
where $\lambda=0$ denotes the trivial isotypic component.

The exact isotypic decomposition of $\mathcal Z$ then follows.
Inserting the resolution of identity $\sum_{\lambda\in\widehat G}\hat P_\lambda=\mathbf 1$ on both sides of $\hat{\mathcal L}^{\mathrm D}$ in the bilinear form appearing in Eq.~\eqref{eq:Z_from_Z1_methods} gives
\begin{align}
\bbra{\mathbf 1}\hat{\mathcal K}_1\hat{\mathcal L}^{\mathrm D}\hat{\mathcal K}_2\kett{\pi}
&=\sum_{\nu,\lambda\in\widehat G}
\bbra{\mathbf 1}
\hat{\mathcal K}_1
\hat P_\nu
\hat{\mathcal L}^{\mathrm D}
\hat P_\lambda
\hat{\mathcal K}_2
\kett{\pi}
\\
&=\sum_{\lambda\in\widehat G}
\bbra{\mathbf 1}
\hat{\mathcal K}_1
\hat P_\lambda
\hat{\mathcal L}^{\mathrm D}
\hat P_\lambda
\hat{\mathcal K}_2
\kett{\pi},
\label{eq:Z1_isotypic_methods}
\end{align}
where the second line follows from Eq.~\eqref{eq:LD_block_isotypic_methods}.
Substituting this decomposition into Eq.~\eqref{eq:Z_from_Z1_methods} yields
\begin{equation}
\mathcal Z
=
\sum_{\lambda\in\widehat G}\mathcal Z^{(\lambda)},
\quad
\mathcal Z^{(\lambda)}
:=
-8{\rm Re}\left[
\bbra{\mathbf 1}
\hat{\mathcal K}_1
\hat P_\lambda
\hat{\mathcal L}^{\mathrm D}
\hat P_\lambda
\hat{\mathcal K}_2
\kett{\pi}
\right].
\label{eq:Z_isotypic_methods}
\end{equation}
This is the exact symmetry-resolved decomposition stated in the main text.

Using the support sets introduced in the main text,
$S_{\rm R}(\hat{\mathcal K}_2)$ and $S_{\rm L}(\hat{\mathcal K}_1)$ in Eq.~\eqref{eq:support_sets_main},
every sector with $\lambda\notin S_{\rm L}(\hat{\mathcal K}_1)\cap S_{\rm R}(\hat{\mathcal K}_2)$ is excluded identically from Eq.~\eqref{eq:Z_isotypic_methods}.
This is the exact classification of symmetry-allowed sectors.

The symmetry-preserving case is recovered first.
If
\begin{equation}
[\hat{\mathcal K}_i,\hat U_g]=0,
\quad \forall g\in G,
\quad i\in\{1,2\},
\label{eq:K_commute_Ug_methods}
\end{equation}
then each $\hat{\mathcal K}_i$ belongs to the commutant of the group representation and therefore preserves every isotypic component:
\begin{equation}
\hat P_\nu \hat{\mathcal K}_i\hat P_\lambda
=
\delta_{\nu\lambda}\hat P_\lambda \hat{\mathcal K}_i.
\label{eq:K_block_isotypic_methods}
\end{equation}
Since $\kett{\pi}$ is trivial, Eq.~\eqref{eq:K_block_isotypic_methods} implies $\hat{\mathcal K}_2\kett{\pi}\in \mathcal H_0$.
Also, $\bbra{\mathbf 1}\hat{\mathcal K}_1$ has support only in $\mathcal H_0$.
Hence $S_{\rm R}(\hat{\mathcal K}_2)=S_{\rm L}(\hat{\mathcal K}_1)=\{0\}$, so Eq.~\eqref{eq:Z_isotypic_methods} reduces to the single term $\lambda=0$.
Every nontrivial irrep is therefore excluded exactly.

Symmetry-covariant perturbations are treated next.
Let $\{\hat T^{(\mu_1)}_a\}_{a=1}^{d_{\mu_1}}$ be an irreducible multiplet contained in $\hat{\mathcal K}_1$ and $\{\hat S^{(\mu_2)}_b\}_{b=1}^{d_{\mu_2}}$ an irreducible multiplet contained in $\hat{\mathcal K}_2$, transforming as
\begin{align}
\hat U_g \hat T^{(\mu_1)}_a \hat U_g^{-1}
&=
\sum_{c=1}^{d_{\mu_1}} D^{(\mu_1)}_{ca}(g)\hat T^{(\mu_1)}_c,
\label{eq:covariant_multiplet_T_methods}\\
\hat U_g \hat S^{(\mu_2)}_b \hat U_g^{-1}
&=
\sum_{c=1}^{d_{\mu_2}} D^{(\mu_2)}_{cb}(g)\hat S^{(\mu_2)}_c.
\label{eq:covariant_multiplet_S_methods}
\end{align}
Because $\kett{\pi}$ is invariant, we have
\begin{equation}
\hat U_g \hat S^{(\mu_2)}_b \kett{\pi}
=
\sum_{c=1}^{d_{\mu_2}} D^{(\mu_2)}_{cb}(g)\hat S^{(\mu_2)}_c\kett{\pi}.
\label{eq:right_intertwiner_methods}
\end{equation}
Thus the map $e_b\mapsto \hat S^{(\mu_2)}_b\kett{\pi}$ intertwines the irrep $\mu_2$ with the Liouville-space representation, so its image lies entirely in the $\mu_2$-isotypic component:
\begin{equation}
\hat P_\lambda \hat S^{(\mu_2)}_b \kett{\pi} = 0,
\quad
\lambda\not\cong \mu_2.
\label{eq:right_support_mu_methods}
\end{equation}

On the bra side, fix an isotypic component $\mathcal H_\lambda$ and define a linear map
$F^{(\mu_1)}_\lambda:\mathcal H_\lambda\to \mathbb C^{d_{\mu_1}}$ by
\begin{equation}
\bigl[F^{(\mu_1)}_\lambda(X)\bigr]_a
:=
\bbra{\mathbf 1}\hat T^{(\mu_1)}_a \hat P_\lambda \kett{X}.
\label{eq:F_map_methods}
\end{equation}
Using $\bbra{\mathbf 1}\hat U_g=\bbra{\mathbf 1}$ and Eq.~\eqref{eq:covariant_multiplet_T_methods}, we obtain
\begin{equation}
F^{(\mu_1)}_\lambda(\hat U_g X)
=
D^{(\mu_1)}(g^{-1})F^{(\mu_1)}_\lambda(X).
\label{eq:F_intertwiner_methods}
\end{equation}
Therefore $F^{(\mu_1)}_\lambda$ is an intertwiner from the $\lambda$-isotypic component to the contragredient channel $\mu_1^*$.
By Schur's lemma, it vanishes unless $\lambda\cong \mu_1^*$.
Equivalently,
\begin{equation}
\bbra{\mathbf 1}\hat T^{(\mu_1)}_a \hat P_\lambda = 0,
\quad
\lambda\not\cong \mu_1^*.
\label{eq:left_support_mu_methods}
\end{equation}

The exact pairwise selection rule now follows.
Define the channel-pair amplitude of the fixed irreducible pair $(\mu_1,\mu_2)$ in the sector $\lambda$ by
\begin{equation}
\Xi_{\lambda}^{ab}
:=
-4
\bbra{\mathbf 1}
\hat T^{(\mu_1)}_a
\hat P_\lambda
\hat{\mathcal L}^{\mathrm D}
\hat P_\lambda
\hat S^{(\mu_2)}_b
\kett{\pi}.
\label{eq:Xi_irrep_pair_methods}
\end{equation}
By Eqs.~\eqref{eq:right_support_mu_methods} and \eqref{eq:left_support_mu_methods}, one has
\begin{equation}
\Xi_{\lambda}^{ab}=0
\quad\text{unless}\quad
\lambda\cong \mu_2\cong \mu_1^*.
\label{eq:lambda_overlap_methods}
\end{equation}
Hence
\begin{equation}
\Xi_{\lambda}^{ab}\neq 0
\quad\Longrightarrow\quad
\mu_2\cong \mu_1^*
\quad\Longleftrightarrow\quad
\mathbf 1\subset \mu_1\otimes \mu_2.
\label{eq:selection_rule_irreps_methods}
\end{equation}
This is the irreducible-channel selection rule.
For reducible perturbations one decomposes $\hat{\mathcal K}_1$ and $\hat{\mathcal K}_2$ into irreducible channels and sums over all contragrediently matched pairs.
For the fixed irreducible pair $(\mu_1,\mu_2)$, the corresponding contribution to the sector-resolved real quantity is
\begin{equation}
\mathcal Z_{(\mu_1,\mu_2)}^{(\lambda)}
=
2 {\rm Re}\sum_{a=1}^{d_{\mu_1}}\sum_{b=1}^{d_{\mu_2}}
\Xi_{\lambda}^{ab}.
\label{eq:Z_pair_sector_methods}
\end{equation}

Equation~\eqref{eq:reduced_resolvent_methods} further isolates where the remaining dynamical complexity resides.
At the level of a fixed irreducible pair $(\mu_1,\mu_2)$, choose an orthonormal basis $\{\kett{\lambda,m,\beta}\}$ on $\mathcal H_\lambda\cong V_\lambda\otimes M_\lambda$, where $m=1,\dots,d_\lambda$ labels the irrep factor and $\beta=1,\dots,\dim M_\lambda$ labels the multiplicity factor.
Write
\begin{align}
\hat P_\lambda \hat S^{(\mu_2)}_b\kett{\pi}
&=
\sum_{m,\beta} c^{(b)}_{m\beta}\kett{\lambda,m,\beta},
\label{eq:c_coeffs_methods}\\
\bbra{\mathbf 1}\hat T^{(\mu_1)}_a\hat P_\lambda
&=
\sum_{m,\beta} d^{(a)*}_{m\beta}\bbra{\lambda,m,\beta}.
\label{eq:d_coeffs_methods}
\end{align}
Using Eq.~\eqref{eq:reduced_resolvent_methods}, Eq.~\eqref{eq:Xi_irrep_pair_methods} becomes
\begin{equation}
\Xi_{\lambda}^{ab}
=
-4\sum_{m=1}^{d_\lambda}
\sum_{\beta,\beta'=1}^{\dim M_\lambda}
d^{(a)*}_{m\beta}
(R_\lambda)_{\beta\beta'}
c^{(b)}_{m\beta'}.
\label{eq:Z1_reduced_multiplicity_methods}
\end{equation}
Thus the irrep index $m$ is contracted trivially, while the nontrivial resolvent structure for each irreducible channel pair survives only on the multiplicity space through $R_\lambda$.

For the full sector contribution $\mathcal Z^{(\lambda)}$, write instead
\begin{align}
\hat P_\lambda \hat{\mathcal K}_2\kett{\pi}
&=
\sum_{m,\beta} c_{m\beta}\kett{\lambda,m,\beta},
\label{eq:c_coeffs_block_methods}\\
\bbra{\mathbf 1}\hat{\mathcal K}_1\hat P_\lambda
&=
\sum_{m,\beta} d^{*}_{m\beta}\bbra{\lambda,m,\beta}.
\label{eq:d_coeffs_block_methods}
\end{align}
Then Eq.~\eqref{eq:reduced_resolvent_methods} together with Eq.~\eqref{eq:Z_isotypic_methods} gives
\begin{equation}
\mathcal Z^{(\lambda)}
=
-8{\rm Re}\left[\sum_{m=1}^{d_\lambda}
\sum_{\beta,\beta'=1}^{\dim M_\lambda}
d^{*}_{m\beta}
(R_\lambda)_{\beta\beta'}
c_{m\beta'}\right].
\label{eq:Z_block_master_formula_methods}
\end{equation}
Thus symmetry fixes the allowed isotypic blocks, while the sign within each allowed block is governed by the reduced pairing of the left and right perturbative couplings through the reduced resolvent $R_\lambda$ on the multiplicity space.
This is the precise sense in which symmetry reduces $\mathcal Z$ to reduced resolvents on multiplicity spaces.

A further exact corollary gives a sign criterion whenever a single conjugate relaxation pair exhausts the contribution of a given symmetry-allowed block.
Let $J_\lambda$ denote the decaying modes contained in the $\lambda$-isotypic component, so that
\begin{equation}
\hat P_\lambda\hat{\mathcal L}^{\mathrm D}\hat P_\lambda
=
\sum_{j\in J_\lambda}\frac{1}{\lambda_j}\kettbbra{x_j}{y_j}.
\label{eq:LD_spectral_block_methods}
\end{equation}
Suppose now that only one isotypic component $\lambda$ contributes to Eq.~\eqref{eq:Z_isotypic_methods}, and that within this block only one complex-conjugate pair of decaying eigenvalues
\begin{equation}
\lambda_\pm=-\alpha\pm i\beta,
\quad \alpha>0,
\label{eq:single_pair_eval_methods}
\end{equation}
contributes.
Define the corresponding modal coefficient by
\begin{equation}
c_+
:=
\bbra{\mathbf 1}\hat{\mathcal K}_1\kettbbra{x_+}{y_+}\hat{\mathcal K}_2\kett{\pi},
\label{eq:cplus_def_methods}
\end{equation}
and choose the conjugate partner such that $\kett{x_-}=J\kett{x_+}$ and $\bbra{y_-}=\bbra{y_+}J$, where $J$ is the antiunitary Hermitian-conjugation involution.
Then $c_-=c_+^*$.
Writing $c_+=|c|e^{i\phi}$, Eq.~\eqref{eq:Z_isotypic_methods} yields
\begin{align}
\mathcal Z^{(\lambda)}
&=
-8{\rm Re}\left(\frac{c_+}{\lambda_+}+\frac{c_+^*}{\lambda_+^*}\right)
= -16{\rm Re}\left(\frac{c_+}{\lambda_+}\right)
\nonumber\\
&=
-\frac{16|c|}{\alpha^2+\beta^2}\bigl(-\alpha\cos\phi+\beta\sin\phi\bigr).
\label{eq:single_pair_negative_Z_methods}
\end{align}
Hence
\begin{equation}
\mathcal Z^{(\lambda)}<0
\ \Longleftrightarrow\ 
{\rm Re}\left(\frac{c_+}{\lambda_+}\right)>0
\ \Longleftrightarrow\ 
-\alpha\cos\phi+\beta\sin\phi>0.
\label{eq:single_pair_negative_Z_criterion_methods}
\end{equation}
The explicit qubit counterexample given above is of this type after the $\mathbb Z_2$ parity decomposition of Liouville space generated by conjugation with $\sigma_x$: only the parity-odd block contributes, and within that block the Liouvillian has a single complex-conjugate relaxation pair.

For an Abelian $U(1)$ symmetry the irreps are one-dimensional charge sectors.
Then $\mu_q^*=\mu_{-q}$, so Eq.~\eqref{eq:selection_rule_irreps_methods} reduces to the charge-selection rule $q_2=-q_1$.
In particular, symmetry-preserving perturbations have $q_1=q_2=0$ and therefore probe only the neutral decaying sector, whereas a perturbation such as $\sigma_x=\sigma_++\sigma_-$ decomposes into charge $+1$ and charge $-1$ channels and can therefore activate the coherence sectors.

\bibliography{ref}

\begin{thebibliography}{55}%
\makeatletter
\providecommand \@ifxundefined [1]{%
 \@ifx{#1\undefined}
}%
\providecommand \@ifnum [1]{%
 \ifnum #1\expandafter \@firstoftwo
 \else \expandafter \@secondoftwo
 \fi
}%
\providecommand \@ifx [1]{%
 \ifx #1\expandafter \@firstoftwo
 \else \expandafter \@secondoftwo
 \fi
}%
\providecommand \natexlab [1]{#1}%
\providecommand \enquote  [1]{``#1''}%
\providecommand \bibnamefont  [1]{#1}%
\providecommand \bibfnamefont [1]{#1}%
\providecommand \citenamefont [1]{#1}%
\providecommand \href@noop [0]{\@secondoftwo}%
\providecommand \href [0]{\begingroup \@sanitize@url \@href}%
\providecommand \@href[1]{\@@startlink{#1}\@@href}%
\providecommand \@@href[1]{\endgroup#1\@@endlink}%
\providecommand \@sanitize@url [0]{\catcode `\\12\catcode `\$12\catcode `\&12\catcode `\#12\catcode `\^12\catcode `\_12\catcode `\%12\relax}%
\providecommand \@@startlink[1]{}%
\providecommand \@@endlink[0]{}%
\providecommand \url  [0]{\begingroup\@sanitize@url \@url }%
\providecommand \@url [1]{\endgroup\@href {#1}{\urlprefix }}%
\providecommand \urlprefix  [0]{URL }%
\providecommand \Eprint [0]{\href }%
\providecommand \doibase [0]{https://doi.org/}%
\providecommand \selectlanguage [0]{\@gobble}%
\providecommand \bibinfo  [0]{\@secondoftwo}%
\providecommand \bibfield  [0]{\@secondoftwo}%
\providecommand \translation [1]{[#1]}%
\providecommand \BibitemOpen [0]{}%
\providecommand \bibitemStop [0]{}%
\providecommand \bibitemNoStop [0]{.\EOS\space}%
\providecommand \EOS [0]{\spacefactor3000\relax}%
\providecommand \BibitemShut  [1]{\csname bibitem#1\endcsname}%
\let\auto@bib@innerbib\@empty
\bibitem [{\citenamefont {Barato}\ and\ \citenamefont {Seifert}(2015)}]{barato2015}%
  \BibitemOpen
  \bibfield  {author} {\bibinfo {author} {\bibfnamefont {A.~C.}\ \bibnamefont {Barato}}\ and\ \bibinfo {author} {\bibfnamefont {U.}~\bibnamefont {Seifert}},\ }\bibfield  {title} {\bibinfo {title} {Thermodynamic {{Uncertainty Relation}} for {{Biomolecular Processes}}},\ }\href {https://doi.org/10.1103/PhysRevLett.114.158101} {\bibfield  {journal} {\bibinfo  {journal} {Phys. Rev. Lett.}\ }\textbf {\bibinfo {volume} {114}},\ \bibinfo {pages} {158101} (\bibinfo {year} {2015})}\BibitemShut {NoStop}%
\bibitem [{\citenamefont {Horowitz}\ and\ \citenamefont {Gingrich}(2020)}]{horowitz2020}%
  \BibitemOpen
  \bibfield  {author} {\bibinfo {author} {\bibfnamefont {J.~M.}\ \bibnamefont {Horowitz}}\ and\ \bibinfo {author} {\bibfnamefont {T.~R.}\ \bibnamefont {Gingrich}},\ }\bibfield  {title} {\bibinfo {title} {Thermodynamic uncertainty relations constrain non-equilibrium fluctuations},\ }\href {https://doi.org/10.1038/s41567-019-0702-6} {\bibfield  {journal} {\bibinfo  {journal} {Nat. Phys.}\ }\textbf {\bibinfo {volume} {16}},\ \bibinfo {pages} {15} (\bibinfo {year} {2020})}\BibitemShut {NoStop}%
\bibitem [{\citenamefont {Liu}\ \emph {et~al.}(2020)\citenamefont {Liu}, \citenamefont {Gong},\ and\ \citenamefont {Ueda}}]{liu2020tur}%
  \BibitemOpen
  \bibfield  {author} {\bibinfo {author} {\bibfnamefont {K.}~\bibnamefont {Liu}}, \bibinfo {author} {\bibfnamefont {Z.}~\bibnamefont {Gong}},\ and\ \bibinfo {author} {\bibfnamefont {M.}~\bibnamefont {Ueda}},\ }\bibfield  {title} {\bibinfo {title} {Thermodynamic uncertainty relation for arbitrary initial states},\ }\href {https://doi.org/10.1103/PhysRevLett.125.140602} {\bibfield  {journal} {\bibinfo  {journal} {Phys. Rev. Lett.}\ }\textbf {\bibinfo {volume} {125}},\ \bibinfo {pages} {140602} (\bibinfo {year} {2020})}\BibitemShut {NoStop}%
\bibitem [{\citenamefont {Dieball}\ and\ \citenamefont {Godec}(2023)}]{Dieball2023}%
  \BibitemOpen
  \bibfield  {author} {\bibinfo {author} {\bibfnamefont {C.}~\bibnamefont {Dieball}}\ and\ \bibinfo {author} {\bibfnamefont {A.}~\bibnamefont {Godec}},\ }\bibfield  {title} {\bibinfo {title} {Direct route to thermodynamic uncertainty relations and their saturation},\ }\href {https://doi.org/10.1103/PhysRevLett.130.087101} {\bibfield  {journal} {\bibinfo  {journal} {Phys. Rev. Lett.}\ }\textbf {\bibinfo {volume} {130}},\ \bibinfo {pages} {087101} (\bibinfo {year} {2023})}\BibitemShut {NoStop}%
\bibitem [{\citenamefont {Di~Terlizzi}\ and\ \citenamefont {Baiesi}(2019)}]{diterlizzi2019}%
  \BibitemOpen
  \bibfield  {author} {\bibinfo {author} {\bibfnamefont {I.}~\bibnamefont {Di~Terlizzi}}\ and\ \bibinfo {author} {\bibfnamefont {M.}~\bibnamefont {Baiesi}},\ }\bibfield  {title} {\bibinfo {title} {Kinetic uncertainty relation},\ }\href {https://doi.org/10.1088/1751-8121/aaee34} {\bibfield  {journal} {\bibinfo  {journal} {J. Phys. A: Math. Theor.}\ }\textbf {\bibinfo {volume} {52}},\ \bibinfo {pages} {02LT03} (\bibinfo {year} {2019})}\BibitemShut {NoStop}%
\bibitem [{\citenamefont {Yan}\ \emph {et~al.}(2019)\citenamefont {Yan}, \citenamefont {Hilfinger}, \citenamefont {Vinnicombe},\ and\ \citenamefont {Paulsson}}]{Yan2019}%
  \BibitemOpen
  \bibfield  {author} {\bibinfo {author} {\bibfnamefont {J.}~\bibnamefont {Yan}}, \bibinfo {author} {\bibfnamefont {A.}~\bibnamefont {Hilfinger}}, \bibinfo {author} {\bibfnamefont {G.}~\bibnamefont {Vinnicombe}},\ and\ \bibinfo {author} {\bibfnamefont {J.}~\bibnamefont {Paulsson}},\ }\bibfield  {title} {\bibinfo {title} {Kinetic uncertainty relations for the control of stochastic reaction networks},\ }\href {https://doi.org/10.1103/PhysRevLett.123.108101} {\bibfield  {journal} {\bibinfo  {journal} {Phys. Rev. Lett.}\ }\textbf {\bibinfo {volume} {123}},\ \bibinfo {pages} {108101} (\bibinfo {year} {2019})}\BibitemShut {NoStop}%
\bibitem [{\citenamefont {Macieszczak}(2024)}]{macieszczak2024}%
  \BibitemOpen
  \bibfield  {author} {\bibinfo {author} {\bibfnamefont {K.}~\bibnamefont {Macieszczak}},\ }\href {https://arxiv.org/abs/2407.09839} {\bibinfo {title} {Ultimate kinetic uncertainty relation and optimal performance of stochastic clocks}} (\bibinfo {year} {2024}),\ \Eprint {https://arxiv.org/abs/2407.09839} {arXiv:2407.09839 [cond-mat.stat-mech]} \BibitemShut {NoStop}%
\bibitem [{\citenamefont {Prech}\ \emph {et~al.}(2025{\natexlab{a}})\citenamefont {Prech}, \citenamefont {Landi}, \citenamefont {Meier}, \citenamefont {Nurgalieva}, \citenamefont {Potts}, \citenamefont {Silva},\ and\ \citenamefont {Mitchison}}]{PrechLandi2025}%
  \BibitemOpen
  \bibfield  {author} {\bibinfo {author} {\bibfnamefont {K.}~\bibnamefont {Prech}}, \bibinfo {author} {\bibfnamefont {G.~T.}\ \bibnamefont {Landi}}, \bibinfo {author} {\bibfnamefont {F.}~\bibnamefont {Meier}}, \bibinfo {author} {\bibfnamefont {N.}~\bibnamefont {Nurgalieva}}, \bibinfo {author} {\bibfnamefont {P.~P.}\ \bibnamefont {Potts}}, \bibinfo {author} {\bibfnamefont {R.}~\bibnamefont {Silva}},\ and\ \bibinfo {author} {\bibfnamefont {M.~T.}\ \bibnamefont {Mitchison}},\ }\bibfield  {title} {\bibinfo {title} {Optimal time estimation and the clock uncertainty relation for stochastic processes},\ }\href {https://doi.org/10.1103/rpls-mp8z} {\bibfield  {journal} {\bibinfo  {journal} {Phys. Rev. X}\ }\textbf {\bibinfo {volume} {15}},\ \bibinfo {pages} {031068} (\bibinfo {year} {2025}{\natexlab{a}})}\BibitemShut {NoStop}%
\bibitem [{\citenamefont {Moreira}\ \emph {et~al.}(2025)\citenamefont {Moreira}, \citenamefont {Radaelli}, \citenamefont {Candeloro}, \citenamefont {Binder},\ and\ \citenamefont {Mitchison}}]{Moreira2025}%
  \BibitemOpen
  \bibfield  {author} {\bibinfo {author} {\bibfnamefont {S.~V.}\ \bibnamefont {Moreira}}, \bibinfo {author} {\bibfnamefont {M.}~\bibnamefont {Radaelli}}, \bibinfo {author} {\bibfnamefont {A.}~\bibnamefont {Candeloro}}, \bibinfo {author} {\bibfnamefont {F.~C.}\ \bibnamefont {Binder}},\ and\ \bibinfo {author} {\bibfnamefont {M.~T.}\ \bibnamefont {Mitchison}},\ }\bibfield  {title} {\bibinfo {title} {Precision bounds for multiple currents in open quantum systems},\ }\href {https://doi.org/10.1103/PhysRevE.111.064107} {\bibfield  {journal} {\bibinfo  {journal} {Phys. Rev. E}\ }\textbf {\bibinfo {volume} {111}},\ \bibinfo {pages} {064107} (\bibinfo {year} {2025})}\BibitemShut {NoStop}%
\bibitem [{\citenamefont {Hasegawa}(2020)}]{hasegawa2020}%
  \BibitemOpen
  \bibfield  {author} {\bibinfo {author} {\bibfnamefont {Y.}~\bibnamefont {Hasegawa}},\ }\bibfield  {title} {\bibinfo {title} {Quantum thermodynamic uncertainty relation for continuous measurement},\ }\href {https://doi.org/10.1103/PhysRevLett.125.050601} {\bibfield  {journal} {\bibinfo  {journal} {Phys. Rev. Lett.}\ }\textbf {\bibinfo {volume} {125}},\ \bibinfo {pages} {050601} (\bibinfo {year} {2020})}\BibitemShut {NoStop}%
\bibitem [{\citenamefont {Kewming}\ \emph {et~al.}(2024)\citenamefont {Kewming}, \citenamefont {Kiely}, \citenamefont {Campbell},\ and\ \citenamefont {Landi}}]{Kewming2024}%
  \BibitemOpen
  \bibfield  {author} {\bibinfo {author} {\bibfnamefont {M.~J.}\ \bibnamefont {Kewming}}, \bibinfo {author} {\bibfnamefont {A.}~\bibnamefont {Kiely}}, \bibinfo {author} {\bibfnamefont {S.}~\bibnamefont {Campbell}},\ and\ \bibinfo {author} {\bibfnamefont {G.~T.}\ \bibnamefont {Landi}},\ }\bibfield  {title} {\bibinfo {title} {First passage times for continuous quantum measurement currents},\ }\href {https://doi.org/10.1103/PhysRevA.109.L050202} {\bibfield  {journal} {\bibinfo  {journal} {Phys. Rev. A}\ }\textbf {\bibinfo {volume} {109}},\ \bibinfo {pages} {L050202} (\bibinfo {year} {2024})}\BibitemShut {NoStop}%
\bibitem [{\citenamefont {Prech}\ \emph {et~al.}(2025{\natexlab{b}})\citenamefont {Prech}, \citenamefont {Potts},\ and\ \citenamefont {Landi}}]{prech2024}%
  \BibitemOpen
  \bibfield  {author} {\bibinfo {author} {\bibfnamefont {K.}~\bibnamefont {Prech}}, \bibinfo {author} {\bibfnamefont {P.~P.}\ \bibnamefont {Potts}},\ and\ \bibinfo {author} {\bibfnamefont {G.~T.}\ \bibnamefont {Landi}},\ }\bibfield  {title} {\bibinfo {title} {Role of quantum coherence in kinetic uncertainty relations},\ }\href {https://doi.org/10.1103/PhysRevLett.134.020401} {\bibfield  {journal} {\bibinfo  {journal} {Phys. Rev. Lett.}\ }\textbf {\bibinfo {volume} {134}},\ \bibinfo {pages} {020401} (\bibinfo {year} {2025}{\natexlab{b}})}\BibitemShut {NoStop}%
\bibitem [{\citenamefont {Gong}\ \emph {et~al.}(2016)\citenamefont {Gong}, \citenamefont {Ashida},\ and\ \citenamefont {Ueda}}]{gong2016}%
  \BibitemOpen
  \bibfield  {author} {\bibinfo {author} {\bibfnamefont {Z.}~\bibnamefont {Gong}}, \bibinfo {author} {\bibfnamefont {Y.}~\bibnamefont {Ashida}},\ and\ \bibinfo {author} {\bibfnamefont {M.}~\bibnamefont {Ueda}},\ }\bibfield  {title} {\bibinfo {title} {Quantum-trajectory thermodynamics with discrete feedback control},\ }\href {https://doi.org/10.1103/PhysRevA.94.012107} {\bibfield  {journal} {\bibinfo  {journal} {Phys. Rev. A}\ }\textbf {\bibinfo {volume} {94}},\ \bibinfo {pages} {012107} (\bibinfo {year} {2016})}\BibitemShut {NoStop}%
\bibitem [{\citenamefont {Liu}\ and\ \citenamefont {Xi}(2016)}]{liuCharacteristic2016}%
  \BibitemOpen
  \bibfield  {author} {\bibinfo {author} {\bibfnamefont {F.}~\bibnamefont {Liu}}\ and\ \bibinfo {author} {\bibfnamefont {J.}~\bibnamefont {Xi}},\ }\bibfield  {title} {\bibinfo {title} {Characteristic functions based on quantum jump trajectory},\ }\href {https://doi.org/10.1103/PhysRevE.94.062133} {\bibfield  {journal} {\bibinfo  {journal} {Phys. Rev. E}\ }\textbf {\bibinfo {volume} {94}},\ \bibinfo {pages} {062133} (\bibinfo {year} {2016})}\BibitemShut {NoStop}%
\bibitem [{\citenamefont {Carollo}\ \emph {et~al.}(2019)\citenamefont {Carollo}, \citenamefont {Jack},\ and\ \citenamefont {Garrahan}}]{carollo2019}%
  \BibitemOpen
  \bibfield  {author} {\bibinfo {author} {\bibfnamefont {F.}~\bibnamefont {Carollo}}, \bibinfo {author} {\bibfnamefont {R.~L.}\ \bibnamefont {Jack}},\ and\ \bibinfo {author} {\bibfnamefont {J.~P.}\ \bibnamefont {Garrahan}},\ }\bibfield  {title} {\bibinfo {title} {Unraveling the {{Large Deviation Statistics}} of {{Markovian Open Quantum Systems}}},\ }\href {https://doi.org/10.1103/PhysRevLett.122.130605} {\bibfield  {journal} {\bibinfo  {journal} {Phys. Rev. Lett.}\ }\textbf {\bibinfo {volume} {122}},\ \bibinfo {pages} {130605} (\bibinfo {year} {2019})}\BibitemShut {NoStop}%
\bibitem [{\citenamefont {Lidar}(2020)}]{lidar2020}%
  \BibitemOpen
  \bibfield  {author} {\bibinfo {author} {\bibfnamefont {D.~A.}\ \bibnamefont {Lidar}},\ }\href@noop {} {\bibinfo {title} {Lecture notes on the theory of open quantum systems}} (\bibinfo {year} {2020}),\ \Eprint {https://arxiv.org/abs/1902.00967} {arXiv:1902.00967 [quant-ph]} \BibitemShut {NoStop}%
\bibitem [{\citenamefont {He}\ \emph {et~al.}(2023)\citenamefont {He}, \citenamefont {Pakkiam}, \citenamefont {Gangat}, \citenamefont {Kewming}, \citenamefont {Milburn},\ and\ \citenamefont {Fedorov}}]{He2023}%
  \BibitemOpen
  \bibfield  {author} {\bibinfo {author} {\bibfnamefont {X.}~\bibnamefont {He}}, \bibinfo {author} {\bibfnamefont {P.}~\bibnamefont {Pakkiam}}, \bibinfo {author} {\bibfnamefont {A.~A.}\ \bibnamefont {Gangat}}, \bibinfo {author} {\bibfnamefont {M.~J.}\ \bibnamefont {Kewming}}, \bibinfo {author} {\bibfnamefont {G.~J.}\ \bibnamefont {Milburn}},\ and\ \bibinfo {author} {\bibfnamefont {A.}~\bibnamefont {Fedorov}},\ }\bibfield  {title} {\bibinfo {title} {Effect of measurement backaction on quantum clock precision studied with a superconducting circuit},\ }\href {https://doi.org/10.1103/PhysRevApplied.20.034038} {\bibfield  {journal} {\bibinfo  {journal} {Phys. Rev. Appl.}\ }\textbf {\bibinfo {volume} {20}},\ \bibinfo {pages} {034038} (\bibinfo {year} {2023})}\BibitemShut {NoStop}%
\bibitem [{\citenamefont {Landi}\ \emph {et~al.}(2024)\citenamefont {Landi}, \citenamefont {Kewming}, \citenamefont {Mitchison},\ and\ \citenamefont {Potts}}]{landi2024}%
  \BibitemOpen
  \bibfield  {author} {\bibinfo {author} {\bibfnamefont {G.~T.}\ \bibnamefont {Landi}}, \bibinfo {author} {\bibfnamefont {M.~J.}\ \bibnamefont {Kewming}}, \bibinfo {author} {\bibfnamefont {M.~T.}\ \bibnamefont {Mitchison}},\ and\ \bibinfo {author} {\bibfnamefont {P.~P.}\ \bibnamefont {Potts}},\ }\bibfield  {title} {\bibinfo {title} {Current fluctuations in open quantum systems: {{Bridging}} the gap between quantum continuous measurements and full counting statistics},\ }\href {https://doi.org/10.1103/PRXQuantum.5.020201} {\bibfield  {journal} {\bibinfo  {journal} {PRX Quantum}\ }\textbf {\bibinfo {volume} {5}},\ \bibinfo {pages} {020201} (\bibinfo {year} {2024})}\BibitemShut {NoStop}%
\bibitem [{\citenamefont {Rossi}\ \emph {et~al.}(2020)\citenamefont {Rossi}, \citenamefont {Albarelli}, \citenamefont {Tamascelli},\ and\ \citenamefont {Genoni}}]{Rossi2020}%
  \BibitemOpen
  \bibfield  {author} {\bibinfo {author} {\bibfnamefont {M.~A.~C.}\ \bibnamefont {Rossi}}, \bibinfo {author} {\bibfnamefont {F.}~\bibnamefont {Albarelli}}, \bibinfo {author} {\bibfnamefont {D.}~\bibnamefont {Tamascelli}},\ and\ \bibinfo {author} {\bibfnamefont {M.~G.}\ \bibnamefont {Genoni}},\ }\bibfield  {title} {\bibinfo {title} {Noisy quantum metrology enhanced by continuous nondemolition measurement},\ }\href {https://doi.org/10.1103/PhysRevLett.125.200505} {\bibfield  {journal} {\bibinfo  {journal} {Phys. Rev. Lett.}\ }\textbf {\bibinfo {volume} {125}},\ \bibinfo {pages} {200505} (\bibinfo {year} {2020})}\BibitemShut {NoStop}%
\bibitem [{\citenamefont {Ilias}\ \emph {et~al.}(2022)\citenamefont {Ilias}, \citenamefont {Yang}, \citenamefont {Huelga},\ and\ \citenamefont {Plenio}}]{Ilias2022}%
  \BibitemOpen
  \bibfield  {author} {\bibinfo {author} {\bibfnamefont {T.}~\bibnamefont {Ilias}}, \bibinfo {author} {\bibfnamefont {D.}~\bibnamefont {Yang}}, \bibinfo {author} {\bibfnamefont {S.~F.}\ \bibnamefont {Huelga}},\ and\ \bibinfo {author} {\bibfnamefont {M.~B.}\ \bibnamefont {Plenio}},\ }\bibfield  {title} {\bibinfo {title} {Criticality-enhanced quantum sensing via continuous measurement},\ }\href {https://doi.org/10.1103/PRXQuantum.3.010354} {\bibfield  {journal} {\bibinfo  {journal} {PRX Quantum}\ }\textbf {\bibinfo {volume} {3}},\ \bibinfo {pages} {010354} (\bibinfo {year} {2022})}\BibitemShut {NoStop}%
\bibitem [{\citenamefont {Yang}\ \emph {et~al.}(2023)\citenamefont {Yang}, \citenamefont {Huelga},\ and\ \citenamefont {Plenio}}]{Yang2023}%
  \BibitemOpen
  \bibfield  {author} {\bibinfo {author} {\bibfnamefont {D.}~\bibnamefont {Yang}}, \bibinfo {author} {\bibfnamefont {S.~F.}\ \bibnamefont {Huelga}},\ and\ \bibinfo {author} {\bibfnamefont {M.~B.}\ \bibnamefont {Plenio}},\ }\bibfield  {title} {\bibinfo {title} {Efficient information retrieval for sensing via continuous measurement},\ }\href {https://doi.org/10.1103/PhysRevX.13.031012} {\bibfield  {journal} {\bibinfo  {journal} {Phys. Rev. X}\ }\textbf {\bibinfo {volume} {13}},\ \bibinfo {pages} {031012} (\bibinfo {year} {2023})}\BibitemShut {NoStop}%
\bibitem [{\citenamefont {Ueda}\ \emph {et~al.}(1990)\citenamefont {Ueda}, \citenamefont {Imoto},\ and\ \citenamefont {Ogawa}}]{ueda1990}%
  \BibitemOpen
  \bibfield  {author} {\bibinfo {author} {\bibfnamefont {M.}~\bibnamefont {Ueda}}, \bibinfo {author} {\bibfnamefont {N.}~\bibnamefont {Imoto}},\ and\ \bibinfo {author} {\bibfnamefont {T.}~\bibnamefont {Ogawa}},\ }\bibfield  {title} {\bibinfo {title} {Quantum theory for continuous photodetection processes},\ }\href {https://doi.org/10.1103/PhysRevA.41.3891} {\bibfield  {journal} {\bibinfo  {journal} {Phys. Rev. A}\ }\textbf {\bibinfo {volume} {41}},\ \bibinfo {pages} {3891} (\bibinfo {year} {1990})}\BibitemShut {NoStop}%
\bibitem [{\citenamefont {Bakr}\ \emph {et~al.}(2009)\citenamefont {Bakr}, \citenamefont {Gillen}, \citenamefont {Peng}, \citenamefont {F{\"o}lling},\ and\ \citenamefont {Greiner}}]{bakr2009}%
  \BibitemOpen
  \bibfield  {author} {\bibinfo {author} {\bibfnamefont {W.~S.}\ \bibnamefont {Bakr}}, \bibinfo {author} {\bibfnamefont {J.~I.}\ \bibnamefont {Gillen}}, \bibinfo {author} {\bibfnamefont {A.}~\bibnamefont {Peng}}, \bibinfo {author} {\bibfnamefont {S.}~\bibnamefont {F{\"o}lling}},\ and\ \bibinfo {author} {\bibfnamefont {M.}~\bibnamefont {Greiner}},\ }\bibfield  {title} {\bibinfo {title} {A quantum gas microscope for detecting single atoms in a hubbard-regime optical lattice},\ }\href {https://doi.org/10.1038/nature08482} {\bibfield  {journal} {\bibinfo  {journal} {Nature (London)}\ }\textbf {\bibinfo {volume} {462}},\ \bibinfo {pages} {74} (\bibinfo {year} {2009})}\BibitemShut {NoStop}%
\bibitem [{\citenamefont {Sherson}\ \emph {et~al.}(2010)\citenamefont {Sherson}, \citenamefont {Weitenberg}, \citenamefont {Endres}, \citenamefont {Cheneau}, \citenamefont {Bloch},\ and\ \citenamefont {Kuhr}}]{sherson2010}%
  \BibitemOpen
  \bibfield  {author} {\bibinfo {author} {\bibfnamefont {J.~F.}\ \bibnamefont {Sherson}}, \bibinfo {author} {\bibfnamefont {C.}~\bibnamefont {Weitenberg}}, \bibinfo {author} {\bibfnamefont {M.}~\bibnamefont {Endres}}, \bibinfo {author} {\bibfnamefont {M.}~\bibnamefont {Cheneau}}, \bibinfo {author} {\bibfnamefont {I.}~\bibnamefont {Bloch}},\ and\ \bibinfo {author} {\bibfnamefont {S.}~\bibnamefont {Kuhr}},\ }\bibfield  {title} {\bibinfo {title} {Single-atom-resolved fluorescence imaging of an atomic mott insulator},\ }\href {https://doi.org/10.1038/nature09378} {\bibfield  {journal} {\bibinfo  {journal} {Nature (London)}\ }\textbf {\bibinfo {volume} {467}},\ \bibinfo {pages} {68} (\bibinfo {year} {2010})}\BibitemShut {NoStop}%
\bibitem [{\citenamefont {Patil}\ \emph {et~al.}(2015)\citenamefont {Patil}, \citenamefont {Chakram},\ and\ \citenamefont {Vengalattore}}]{Patil2015}%
  \BibitemOpen
  \bibfield  {author} {\bibinfo {author} {\bibfnamefont {Y.~S.}\ \bibnamefont {Patil}}, \bibinfo {author} {\bibfnamefont {S.}~\bibnamefont {Chakram}},\ and\ \bibinfo {author} {\bibfnamefont {M.}~\bibnamefont {Vengalattore}},\ }\bibfield  {title} {\bibinfo {title} {Measurement-induced localization of an ultracold lattice gas},\ }\href {https://doi.org/10.1103/PhysRevLett.115.140402} {\bibfield  {journal} {\bibinfo  {journal} {Phys. Rev. Lett.}\ }\textbf {\bibinfo {volume} {115}},\ \bibinfo {pages} {140402} (\bibinfo {year} {2015})}\BibitemShut {NoStop}%
\bibitem [{\citenamefont {Ashida}\ and\ \citenamefont {Ueda}(2015)}]{Ashida2015}%
  \BibitemOpen
  \bibfield  {author} {\bibinfo {author} {\bibfnamefont {Y.}~\bibnamefont {Ashida}}\ and\ \bibinfo {author} {\bibfnamefont {M.}~\bibnamefont {Ueda}},\ }\bibfield  {title} {\bibinfo {title} {Diffraction-unlimited position measurement of ultracold atoms in an optical lattice},\ }\href {https://doi.org/10.1103/PhysRevLett.115.095301} {\bibfield  {journal} {\bibinfo  {journal} {Phys. Rev. Lett.}\ }\textbf {\bibinfo {volume} {115}},\ \bibinfo {pages} {095301} (\bibinfo {year} {2015})}\BibitemShut {NoStop}%
\bibitem [{\citenamefont {Wigley}\ \emph {et~al.}(2016)\citenamefont {Wigley}, \citenamefont {Everitt}, \citenamefont {Hardman}, \citenamefont {Hush}, \citenamefont {Wei}, \citenamefont {Sooriyabandara}, \citenamefont {Manju}, \citenamefont {Close}, \citenamefont {Robins},\ and\ \citenamefont {Kuhn}}]{Wigley2016}%
  \BibitemOpen
  \bibfield  {author} {\bibinfo {author} {\bibfnamefont {P.~B.}\ \bibnamefont {Wigley}}, \bibinfo {author} {\bibfnamefont {P.~J.}\ \bibnamefont {Everitt}}, \bibinfo {author} {\bibfnamefont {K.~S.}\ \bibnamefont {Hardman}}, \bibinfo {author} {\bibfnamefont {M.~R.}\ \bibnamefont {Hush}}, \bibinfo {author} {\bibfnamefont {C.~H.}\ \bibnamefont {Wei}}, \bibinfo {author} {\bibfnamefont {M.~A.}\ \bibnamefont {Sooriyabandara}}, \bibinfo {author} {\bibfnamefont {P.}~\bibnamefont {Manju}}, \bibinfo {author} {\bibfnamefont {J.~D.}\ \bibnamefont {Close}}, \bibinfo {author} {\bibfnamefont {N.~P.}\ \bibnamefont {Robins}},\ and\ \bibinfo {author} {\bibfnamefont {C.~C.~N.}\ \bibnamefont {Kuhn}},\ }\bibfield  {title} {\bibinfo {title} {Non-destructive shadowgraph imaging of ultra-cold atoms},\ }\href {https://doi.org/10.1364/OL.41.004795} {\bibfield  {journal} {\bibinfo  {journal} {Opt. Lett.}\ }\textbf {\bibinfo {volume} {41}},\ \bibinfo {pages} {4795} (\bibinfo {year} {2016})}\BibitemShut {NoStop}%
\bibitem [{\citenamefont {Gross}\ and\ \citenamefont {Bakr}(2021)}]{gross2021}%
  \BibitemOpen
  \bibfield  {author} {\bibinfo {author} {\bibfnamefont {C.}~\bibnamefont {Gross}}\ and\ \bibinfo {author} {\bibfnamefont {W.~S.}\ \bibnamefont {Bakr}},\ }\bibfield  {title} {\bibinfo {title} {Quantum gas microscopy for single atom and spin detection},\ }\href {https://doi.org/10.1038/s41567-021-01370-5} {\bibfield  {journal} {\bibinfo  {journal} {Nat. Phys.}\ }\textbf {\bibinfo {volume} {17}},\ \bibinfo {pages} {1316} (\bibinfo {year} {2021})}\BibitemShut {NoStop}%
\bibitem [{\citenamefont {Fuji}\ and\ \citenamefont {Ashida}(2020)}]{Fuji2020}%
  \BibitemOpen
  \bibfield  {author} {\bibinfo {author} {\bibfnamefont {Y.}~\bibnamefont {Fuji}}\ and\ \bibinfo {author} {\bibfnamefont {Y.}~\bibnamefont {Ashida}},\ }\bibfield  {title} {\bibinfo {title} {Measurement-induced quantum criticality under continuous monitoring},\ }\href {https://doi.org/10.1103/PhysRevB.102.054302} {\bibfield  {journal} {\bibinfo  {journal} {Phys. Rev. B}\ }\textbf {\bibinfo {volume} {102}},\ \bibinfo {pages} {054302} (\bibinfo {year} {2020})}\BibitemShut {NoStop}%
\bibitem [{\citenamefont {Jian}\ \emph {et~al.}(2021)\citenamefont {Jian}, \citenamefont {Liu}, \citenamefont {Chen}, \citenamefont {Swingle},\ and\ \citenamefont {Zhang}}]{Jian2021}%
  \BibitemOpen
  \bibfield  {author} {\bibinfo {author} {\bibfnamefont {S.-K.}\ \bibnamefont {Jian}}, \bibinfo {author} {\bibfnamefont {C.}~\bibnamefont {Liu}}, \bibinfo {author} {\bibfnamefont {X.}~\bibnamefont {Chen}}, \bibinfo {author} {\bibfnamefont {B.}~\bibnamefont {Swingle}},\ and\ \bibinfo {author} {\bibfnamefont {P.}~\bibnamefont {Zhang}},\ }\bibfield  {title} {\bibinfo {title} {Measurement-induced phase transition in the monitored sachdev-ye-kitaev model},\ }\href {https://doi.org/10.1103/PhysRevLett.127.140601} {\bibfield  {journal} {\bibinfo  {journal} {Phys. Rev. Lett.}\ }\textbf {\bibinfo {volume} {127}},\ \bibinfo {pages} {140601} (\bibinfo {year} {2021})}\BibitemShut {NoStop}%
\bibitem [{\citenamefont {M\"uller}\ \emph {et~al.}(2022)\citenamefont {M\"uller}, \citenamefont {Diehl},\ and\ \citenamefont {Buchhold}}]{muller2022}%
  \BibitemOpen
  \bibfield  {author} {\bibinfo {author} {\bibfnamefont {T.}~\bibnamefont {M\"uller}}, \bibinfo {author} {\bibfnamefont {S.}~\bibnamefont {Diehl}},\ and\ \bibinfo {author} {\bibfnamefont {M.}~\bibnamefont {Buchhold}},\ }\bibfield  {title} {\bibinfo {title} {Measurement-induced dark state phase transitions in long-ranged fermion systems},\ }\href {https://doi.org/10.1103/PhysRevLett.128.010605} {\bibfield  {journal} {\bibinfo  {journal} {Phys. Rev. Lett.}\ }\textbf {\bibinfo {volume} {128}},\ \bibinfo {pages} {010605} (\bibinfo {year} {2022})}\BibitemShut {NoStop}%
\bibitem [{\citenamefont {Krishna}\ \emph {et~al.}(2023)\citenamefont {Krishna}, \citenamefont {Solanki}, \citenamefont {Hajdu\ifmmode~\check{s}\else \v{s}\fi{}ek},\ and\ \citenamefont {Vinjanampathy}}]{Krishna2023}%
  \BibitemOpen
  \bibfield  {author} {\bibinfo {author} {\bibfnamefont {M.}~\bibnamefont {Krishna}}, \bibinfo {author} {\bibfnamefont {P.}~\bibnamefont {Solanki}}, \bibinfo {author} {\bibfnamefont {M.}~\bibnamefont {Hajdu\ifmmode~\check{s}\else \v{s}\fi{}ek}},\ and\ \bibinfo {author} {\bibfnamefont {S.}~\bibnamefont {Vinjanampathy}},\ }\bibfield  {title} {\bibinfo {title} {Measurement-induced continuous time crystals},\ }\href {https://doi.org/10.1103/PhysRevLett.130.150401} {\bibfield  {journal} {\bibinfo  {journal} {Phys. Rev. Lett.}\ }\textbf {\bibinfo {volume} {130}},\ \bibinfo {pages} {150401} (\bibinfo {year} {2023})}\BibitemShut {NoStop}%
\bibitem [{\citenamefont {Yokomizo}\ and\ \citenamefont {Ashida}(2025)}]{yokomizo2024}%
  \BibitemOpen
  \bibfield  {author} {\bibinfo {author} {\bibfnamefont {K.}~\bibnamefont {Yokomizo}}\ and\ \bibinfo {author} {\bibfnamefont {Y.}~\bibnamefont {Ashida}},\ }\bibfield  {title} {\bibinfo {title} {Measurement-induced phase transition in free bosons},\ }\href {https://doi.org/10.1103/y5r3-tv78} {\bibfield  {journal} {\bibinfo  {journal} {Phys. Rev. B}\ }\textbf {\bibinfo {volume} {111}},\ \bibinfo {pages} {235419} (\bibinfo {year} {2025})}\BibitemShut {NoStop}%
\bibitem [{\citenamefont {Ashida}\ \emph {et~al.}(2018)\citenamefont {Ashida}, \citenamefont {Saito},\ and\ \citenamefont {Ueda}}]{Ashida2018}%
  \BibitemOpen
  \bibfield  {author} {\bibinfo {author} {\bibfnamefont {Y.}~\bibnamefont {Ashida}}, \bibinfo {author} {\bibfnamefont {K.}~\bibnamefont {Saito}},\ and\ \bibinfo {author} {\bibfnamefont {M.}~\bibnamefont {Ueda}},\ }\bibfield  {title} {\bibinfo {title} {Thermalization and heating dynamics in open generic many-body systems},\ }\href {https://doi.org/10.1103/PhysRevLett.121.170402} {\bibfield  {journal} {\bibinfo  {journal} {Phys. Rev. Lett.}\ }\textbf {\bibinfo {volume} {121}},\ \bibinfo {pages} {170402} (\bibinfo {year} {2018})}\BibitemShut {NoStop}%
\bibitem [{\citenamefont {Kewming}\ and\ \citenamefont {Shrapnel}(2022)}]{Kewming2022entropyproduction}%
  \BibitemOpen
  \bibfield  {author} {\bibinfo {author} {\bibfnamefont {M.~J.}\ \bibnamefont {Kewming}}\ and\ \bibinfo {author} {\bibfnamefont {S.}~\bibnamefont {Shrapnel}},\ }\bibfield  {title} {\bibinfo {title} {Entropy production and fluctuation theorems in a continuously monitored optical cavity at zero temperature},\ }\href {https://doi.org/10.22331/q-2022-04-13-685} {\bibfield  {journal} {\bibinfo  {journal} {{Quantum}}\ }\textbf {\bibinfo {volume} {6}},\ \bibinfo {pages} {685} (\bibinfo {year} {2022})}\BibitemShut {NoStop}%
\bibitem [{\citenamefont {Manzano}\ and\ \citenamefont {Zambrini}(2022)}]{Manzano2022}%
  \BibitemOpen
  \bibfield  {author} {\bibinfo {author} {\bibfnamefont {G.}~\bibnamefont {Manzano}}\ and\ \bibinfo {author} {\bibfnamefont {R.}~\bibnamefont {Zambrini}},\ }\bibfield  {title} {\bibinfo {title} {Quantum thermodynamics under continuous monitoring: A general framework},\ }\href {https://doi.org/10.1116/5.0079886} {\bibfield  {journal} {\bibinfo  {journal} {AVS Quantum Sci.}\ }\textbf {\bibinfo {volume} {4}},\ \bibinfo {pages} {025302} (\bibinfo {year} {2022})}\BibitemShut {NoStop}%
\bibitem [{\citenamefont {Das}\ \emph {et~al.}(2023)\citenamefont {Das}, \citenamefont {Mahunta}, \citenamefont {Agarwalla},\ and\ \citenamefont {Mukherjee}}]{Das2023}%
  \BibitemOpen
  \bibfield  {author} {\bibinfo {author} {\bibfnamefont {A.}~\bibnamefont {Das}}, \bibinfo {author} {\bibfnamefont {S.}~\bibnamefont {Mahunta}}, \bibinfo {author} {\bibfnamefont {B.~K.}\ \bibnamefont {Agarwalla}},\ and\ \bibinfo {author} {\bibfnamefont {V.}~\bibnamefont {Mukherjee}},\ }\bibfield  {title} {\bibinfo {title} {Precision bound and optimal control in periodically modulated continuous quantum thermal machines},\ }\href {https://doi.org/10.1103/PhysRevE.108.014137} {\bibfield  {journal} {\bibinfo  {journal} {Phys. Rev. E}\ }\textbf {\bibinfo {volume} {108}},\ \bibinfo {pages} {014137} (\bibinfo {year} {2023})}\BibitemShut {NoStop}%
\bibitem [{\citenamefont {Ptaszy\ifmmode~\acute{n}\else \'{n}\fi{}ski}\ \emph {et~al.}(2024)\citenamefont {Ptaszy\ifmmode~\acute{n}\else \'{n}\fi{}ski}, \citenamefont {Aslyamov},\ and\ \citenamefont {Esposito}}]{ptaszynski2024}%
  \BibitemOpen
  \bibfield  {author} {\bibinfo {author} {\bibfnamefont {K.}~\bibnamefont {Ptaszy\ifmmode~\acute{n}\else \'{n}\fi{}ski}}, \bibinfo {author} {\bibfnamefont {T.}~\bibnamefont {Aslyamov}},\ and\ \bibinfo {author} {\bibfnamefont {M.}~\bibnamefont {Esposito}},\ }\bibfield  {title} {\bibinfo {title} {Dissipation bounds precision of current response to kinetic perturbations},\ }\href {https://doi.org/10.1103/PhysRevLett.133.227101} {\bibfield  {journal} {\bibinfo  {journal} {Phys. Rev. Lett.}\ }\textbf {\bibinfo {volume} {133}},\ \bibinfo {pages} {227101} (\bibinfo {year} {2024})}\BibitemShut {NoStop}%
\bibitem [{\citenamefont {Liu}\ and\ \citenamefont {Gu}(2025)}]{liu2024}%
  \BibitemOpen
  \bibfield  {author} {\bibinfo {author} {\bibfnamefont {K.}~\bibnamefont {Liu}}\ and\ \bibinfo {author} {\bibfnamefont {J.}~\bibnamefont {Gu}},\ }\bibfield  {title} {\bibinfo {title} {Dynamical activity universally bounds precision of response in markovian nonequilibrium systems},\ }\href {https://doi.org/10.1038/s42005-025-01982-w} {\bibfield  {journal} {\bibinfo  {journal} {Commun. Phys.}\ }\textbf {\bibinfo {volume} {8}},\ \bibinfo {pages} {62} (\bibinfo {year} {2025})}\BibitemShut {NoStop}%
\bibitem [{\citenamefont {Aslyamov}\ \emph {et~al.}(2025)\citenamefont {Aslyamov}, \citenamefont {Ptaszy\ifmmode~\acute{n}\else \'{n}\fi{}ski},\ and\ \citenamefont {Esposito}}]{aslyamov2024}%
  \BibitemOpen
  \bibfield  {author} {\bibinfo {author} {\bibfnamefont {T.}~\bibnamefont {Aslyamov}}, \bibinfo {author} {\bibfnamefont {K.}~\bibnamefont {Ptaszy\ifmmode~\acute{n}\else \'{n}\fi{}ski}},\ and\ \bibinfo {author} {\bibfnamefont {M.}~\bibnamefont {Esposito}},\ }\bibfield  {title} {\bibinfo {title} {Nonequilibrium fluctuation-response relations: From identities to bounds},\ }\href {https://doi.org/10.1103/PhysRevLett.134.157101} {\bibfield  {journal} {\bibinfo  {journal} {Phys. Rev. Lett.}\ }\textbf {\bibinfo {volume} {134}},\ \bibinfo {pages} {157101} (\bibinfo {year} {2025})}\BibitemShut {NoStop}%
\bibitem [{\citenamefont {Kwon}\ \emph {et~al.}(2025)\citenamefont {Kwon}, \citenamefont {Chun}, \citenamefont {Park},\ and\ \citenamefont {Lee}}]{kwon2024fri}%
  \BibitemOpen
  \bibfield  {author} {\bibinfo {author} {\bibfnamefont {E.}~\bibnamefont {Kwon}}, \bibinfo {author} {\bibfnamefont {H.-M.}\ \bibnamefont {Chun}}, \bibinfo {author} {\bibfnamefont {H.}~\bibnamefont {Park}},\ and\ \bibinfo {author} {\bibfnamefont {J.~S.}\ \bibnamefont {Lee}},\ }\bibfield  {title} {\bibinfo {title} {Fluctuation-response inequalities for kinetic and entropic perturbations},\ }\href {https://doi.org/10.1103/PhysRevLett.135.097101} {\bibfield  {journal} {\bibinfo  {journal} {Phys. Rev. Lett.}\ }\textbf {\bibinfo {volume} {135}},\ \bibinfo {pages} {097101} (\bibinfo {year} {2025})}\BibitemShut {NoStop}%
\bibitem [{\citenamefont {Nakajima}\ and\ \citenamefont {Utsumi}(2023)}]{NakajimaUtsumi2023}%
  \BibitemOpen
  \bibfield  {author} {\bibinfo {author} {\bibfnamefont {S.}~\bibnamefont {Nakajima}}\ and\ \bibinfo {author} {\bibfnamefont {Y.}~\bibnamefont {Utsumi}},\ }\bibfield  {title} {\bibinfo {title} {Symmetric-logarithmic-derivative fisher information for kinetic uncertainty relations},\ }\href {https://doi.org/10.1103/PhysRevE.108.054136} {\bibfield  {journal} {\bibinfo  {journal} {Phys. Rev. E}\ }\textbf {\bibinfo {volume} {108}},\ \bibinfo {pages} {054136} (\bibinfo {year} {2023})}\BibitemShut {NoStop}%
\bibitem [{\citenamefont {Van~Vu}(2025)}]{vanvu2024}%
  \BibitemOpen
  \bibfield  {author} {\bibinfo {author} {\bibfnamefont {T.}~\bibnamefont {Van~Vu}},\ }\bibfield  {title} {\bibinfo {title} {Fundamental bounds on precision and response for quantum trajectory observables},\ }\href {https://doi.org/10.1103/PRXQuantum.6.010343} {\bibfield  {journal} {\bibinfo  {journal} {PRX Quantum}\ }\textbf {\bibinfo {volume} {6}},\ \bibinfo {pages} {010343} (\bibinfo {year} {2025})}\BibitemShut {NoStop}%
\bibitem [{\citenamefont {Lindblad}(1976)}]{Lindblad1976}%
  \BibitemOpen
  \bibfield  {author} {\bibinfo {author} {\bibfnamefont {G.}~\bibnamefont {Lindblad}},\ }\bibfield  {title} {\bibinfo {title} {On the generators of quantum dynamical semigroups},\ }\href {https://doi.org/10.1007/BF01608499} {\bibfield  {journal} {\bibinfo  {journal} {Commun. Math. Phys.}\ }\textbf {\bibinfo {volume} {48}},\ \bibinfo {pages} {119} (\bibinfo {year} {1976})}\BibitemShut {NoStop}%
\bibitem [{\citenamefont {Boeyens}\ \emph {et~al.}(2023)\citenamefont {Boeyens}, \citenamefont {{Annby-Andersson}}, \citenamefont {Bakhshinezhad}, \citenamefont {Haack}, \citenamefont {{Perarnau-Llobet}}, \citenamefont {Nimmrichter}, \citenamefont {Potts},\ and\ \citenamefont {Mehboudi}}]{boeyens2023}%
  \BibitemOpen
  \bibfield  {author} {\bibinfo {author} {\bibfnamefont {J.}~\bibnamefont {Boeyens}}, \bibinfo {author} {\bibfnamefont {B.}~\bibnamefont {{Annby-Andersson}}}, \bibinfo {author} {\bibfnamefont {P.}~\bibnamefont {Bakhshinezhad}}, \bibinfo {author} {\bibfnamefont {G.}~\bibnamefont {Haack}}, \bibinfo {author} {\bibfnamefont {M.}~\bibnamefont {{Perarnau-Llobet}}}, \bibinfo {author} {\bibfnamefont {S.}~\bibnamefont {Nimmrichter}}, \bibinfo {author} {\bibfnamefont {P.~P.}\ \bibnamefont {Potts}},\ and\ \bibinfo {author} {\bibfnamefont {M.}~\bibnamefont {Mehboudi}},\ }\bibfield  {title} {\bibinfo {title} {Probe thermometry with continuous measurements},\ }\href {https://doi.org/10.1088/1367-2630/ad0e8a} {\bibfield  {journal} {\bibinfo  {journal} {New J. Phys.}\ }\textbf {\bibinfo {volume} {25}},\ \bibinfo {pages} {123009} (\bibinfo {year} {2023})}\BibitemShut {NoStop}%
\bibitem [{\citenamefont {Gammelmark}\ and\ \citenamefont {M{\o}lmer}(2014)}]{gammelmark2014}%
  \BibitemOpen
  \bibfield  {author} {\bibinfo {author} {\bibfnamefont {S.}~\bibnamefont {Gammelmark}}\ and\ \bibinfo {author} {\bibfnamefont {K.}~\bibnamefont {M{\o}lmer}},\ }\bibfield  {title} {\bibinfo {title} {The {{Fisher}} information and the quantum {{Cramer-Rao}} sensitivity limit of continuous measurements},\ }\href {https://doi.org/10.1103/PhysRevLett.112.170401} {\bibfield  {journal} {\bibinfo  {journal} {Phys. Rev. Lett.}\ }\textbf {\bibinfo {volume} {112}},\ \bibinfo {pages} {170401} (\bibinfo {year} {2014})}\BibitemShut {NoStop}%
\bibitem [{\citenamefont {Dalibard}\ \emph {et~al.}(1992)\citenamefont {Dalibard}, \citenamefont {Castin},\ and\ \citenamefont {M\o{}lmer}}]{DalibardCastinMolmer1992}%
  \BibitemOpen
  \bibfield  {author} {\bibinfo {author} {\bibfnamefont {J.}~\bibnamefont {Dalibard}}, \bibinfo {author} {\bibfnamefont {Y.}~\bibnamefont {Castin}},\ and\ \bibinfo {author} {\bibfnamefont {K.}~\bibnamefont {M\o{}lmer}},\ }\bibfield  {title} {\bibinfo {title} {Wave-function approach to dissipative processes in quantum optics},\ }\href {https://doi.org/10.1103/PhysRevLett.68.580} {\bibfield  {journal} {\bibinfo  {journal} {Phys. Rev. Lett.}\ }\textbf {\bibinfo {volume} {68}},\ \bibinfo {pages} {580} (\bibinfo {year} {1992})}\BibitemShut {NoStop}%
\bibitem [{\citenamefont {Braunstein}\ and\ \citenamefont {Caves}(1994)}]{BraunsteinCaves1994}%
  \BibitemOpen
  \bibfield  {author} {\bibinfo {author} {\bibfnamefont {S.~L.}\ \bibnamefont {Braunstein}}\ and\ \bibinfo {author} {\bibfnamefont {C.~M.}\ \bibnamefont {Caves}},\ }\bibfield  {title} {\bibinfo {title} {Statistical distance and the geometry of quantum states},\ }\href {https://doi.org/10.1103/PhysRevLett.72.3439} {\bibfield  {journal} {\bibinfo  {journal} {Phys. Rev. Lett.}\ }\textbf {\bibinfo {volume} {72}},\ \bibinfo {pages} {3439} (\bibinfo {year} {1994})}\BibitemShut {NoStop}%
\bibitem [{\citenamefont {Paris}(2009)}]{Paris2009QuantumEstimation}%
  \BibitemOpen
  \bibfield  {author} {\bibinfo {author} {\bibfnamefont {M.~G.~A.}\ \bibnamefont {Paris}},\ }\bibfield  {title} {\bibinfo {title} {Quantum estimation for quantum technology},\ }\href {https://doi.org/10.1142/S0219749909004839} {\bibfield  {journal} {\bibinfo  {journal} {Int. J. Quantum Inf.}\ }\textbf {\bibinfo {volume} {7}},\ \bibinfo {pages} {125} (\bibinfo {year} {2009})}\BibitemShut {NoStop}%
\bibitem [{\citenamefont {Samach}\ \emph {et~al.}(2022)\citenamefont {Samach}, \citenamefont {Greene}, \citenamefont {Borregaard}, \citenamefont {Christandl}, \citenamefont {Barreto}, \citenamefont {Kim}, \citenamefont {McNally}, \citenamefont {Melville}, \citenamefont {Niedzielski}, \citenamefont {Sung}, \citenamefont {Rosenberg}, \citenamefont {Schwartz}, \citenamefont {Yoder}, \citenamefont {Orlando}, \citenamefont {Wang}, \citenamefont {Gustavsson}, \citenamefont {Kjaergaard},\ and\ \citenamefont {Oliver}}]{PhysRevApplied.18.064056}%
  \BibitemOpen
  \bibfield  {author} {\bibinfo {author} {\bibfnamefont {G.~O.}\ \bibnamefont {Samach}}, \bibinfo {author} {\bibfnamefont {A.}~\bibnamefont {Greene}}, \bibinfo {author} {\bibfnamefont {J.}~\bibnamefont {Borregaard}}, \bibinfo {author} {\bibfnamefont {M.}~\bibnamefont {Christandl}}, \bibinfo {author} {\bibfnamefont {J.}~\bibnamefont {Barreto}}, \bibinfo {author} {\bibfnamefont {D.~K.}\ \bibnamefont {Kim}}, \bibinfo {author} {\bibfnamefont {C.~M.}\ \bibnamefont {McNally}}, \bibinfo {author} {\bibfnamefont {A.}~\bibnamefont {Melville}}, \bibinfo {author} {\bibfnamefont {B.~M.}\ \bibnamefont {Niedzielski}}, \bibinfo {author} {\bibfnamefont {Y.}~\bibnamefont {Sung}}, \bibinfo {author} {\bibfnamefont {D.}~\bibnamefont {Rosenberg}}, \bibinfo {author} {\bibfnamefont {M.~E.}\ \bibnamefont {Schwartz}}, \bibinfo {author} {\bibfnamefont {J.~L.}\ \bibnamefont {Yoder}}, \bibinfo {author} {\bibfnamefont {T.~P.}\ \bibnamefont {Orlando}}, \bibinfo {author} {\bibfnamefont {J.~I.-J.}\ \bibnamefont {Wang}}, \bibinfo {author} {\bibfnamefont {S.}~\bibnamefont {Gustavsson}}, \bibinfo {author} {\bibfnamefont {M.}~\bibnamefont {Kjaergaard}},\ and\ \bibinfo {author} {\bibfnamefont {W.~D.}\ \bibnamefont {Oliver}},\ }\bibfield  {title} {\bibinfo {title} {Lindblad tomography of a superconducting quantum processor},\ }\href {https://doi.org/10.1103/PhysRevApplied.18.064056} {\bibfield  {journal} {\bibinfo  {journal} {Phys. Rev. Appl.}\ }\textbf {\bibinfo {volume} {18}},\ \bibinfo {pages} {064056} (\bibinfo {year} {2022})}\BibitemShut {NoStop}%
\bibitem [{\citenamefont {Onorati}\ \emph {et~al.}(2023)\citenamefont {Onorati}, \citenamefont {Kohler},\ and\ \citenamefont {Cubitt}}]{Onorati2023fittingquantumnoise}%
  \BibitemOpen
  \bibfield  {author} {\bibinfo {author} {\bibfnamefont {E.}~\bibnamefont {Onorati}}, \bibinfo {author} {\bibfnamefont {T.}~\bibnamefont {Kohler}},\ and\ \bibinfo {author} {\bibfnamefont {T.~S.}\ \bibnamefont {Cubitt}},\ }\bibfield  {title} {\bibinfo {title} {Fitting quantum noise models to tomography data},\ }\href {https://doi.org/10.22331/q-2023-12-05-1197} {\bibfield  {journal} {\bibinfo  {journal} {{Quantum}}\ }\textbf {\bibinfo {volume} {7}},\ \bibinfo {pages} {1197} (\bibinfo {year} {2023})}\BibitemShut {NoStop}%
\bibitem [{\citenamefont {Bu{\v{c}}a}\ and\ \citenamefont {Prosen}(2012)}]{BucaProsen2012}%
  \BibitemOpen
  \bibfield  {author} {\bibinfo {author} {\bibfnamefont {B.}~\bibnamefont {Bu{\v{c}}a}}\ and\ \bibinfo {author} {\bibfnamefont {T.}~\bibnamefont {Prosen}},\ }\bibfield  {title} {\bibinfo {title} {A note on symmetry reductions of the lindblad equation: transport in constrained open spin chains},\ }\href {https://doi.org/10.1088/1367-2630/14/7/073007} {\bibfield  {journal} {\bibinfo  {journal} {New J. Phys.}\ }\textbf {\bibinfo {volume} {14}},\ \bibinfo {pages} {073007} (\bibinfo {year} {2012})}\BibitemShut {NoStop}%
\bibitem [{\citenamefont {Albert}\ and\ \citenamefont {Jiang}(2014)}]{AlbertJiang2014}%
  \BibitemOpen
  \bibfield  {author} {\bibinfo {author} {\bibfnamefont {V.~V.}\ \bibnamefont {Albert}}\ and\ \bibinfo {author} {\bibfnamefont {L.}~\bibnamefont {Jiang}},\ }\bibfield  {title} {\bibinfo {title} {Symmetries and conserved quantities in lindblad master equations},\ }\href {https://doi.org/10.1103/PhysRevA.89.022118} {\bibfield  {journal} {\bibinfo  {journal} {Phys. Rev. A}\ }\textbf {\bibinfo {volume} {89}},\ \bibinfo {pages} {022118} (\bibinfo {year} {2014})}\BibitemShut {NoStop}%
\bibitem [{\citenamefont {Latune}\ \emph {et~al.}(2019)\citenamefont {Latune}, \citenamefont {Sinayskiy},\ and\ \citenamefont {Petruccione}}]{Latune2019}%
  \BibitemOpen
  \bibfield  {author} {\bibinfo {author} {\bibfnamefont {C.~L.}\ \bibnamefont {Latune}}, \bibinfo {author} {\bibfnamefont {I.}~\bibnamefont {Sinayskiy}},\ and\ \bibinfo {author} {\bibfnamefont {F.}~\bibnamefont {Petruccione}},\ }\bibfield  {title} {\bibinfo {title} {Energetic and entropic effects of bath-induced coherences},\ }\href {https://doi.org/10.1103/PhysRevA.99.052105} {\bibfield  {journal} {\bibinfo  {journal} {Phys. Rev. A}\ }\textbf {\bibinfo {volume} {99}},\ \bibinfo {pages} {052105} (\bibinfo {year} {2019})}\BibitemShut {NoStop}%
\bibitem [{\citenamefont {Drazin}(1958)}]{Drazin1958}%
  \BibitemOpen
  \bibfield  {author} {\bibinfo {author} {\bibfnamefont {M.~P.}\ \bibnamefont {Drazin}},\ }\bibfield  {title} {\bibinfo {title} {Pseudo-inverses in associative rings and semigroups},\ }\href {http://www.jstor.org/stable/2308576} {\bibfield  {journal} {\bibinfo  {journal} {Am. Math. Mon.}\ }\textbf {\bibinfo {volume} {65}},\ \bibinfo {pages} {506} (\bibinfo {year} {1958})}\BibitemShut {NoStop}%
\end{thebibliography}%

\end{document}